%% file: main.tex
\documentclass[sigconf]{acmart}
\usepackage{multirow}
\usepackage{soul}

\AtBeginDocument{%
  }

\newcommand{\ladd}[1]{#1}
\newcommand{\lrem}[1]{}

\copyrightyear{2025}
\acmYear{2025}
\setcopyright{cc}
\setcctype{by}
\acmConference[CHI '25]{CHI Conference on Human Factors in Computing Systems}{April 26-May 1, 2025}{Yokohama, Japan}
\acmBooktitle{CHI Conference on Human Factors in Computing Systems (CHI '25), April 26-May 1, 2025, Yokohama, Japan}\acmDOI{10.1145/3706598.3713508}
\acmISBN{979-8-4007-1394-1/25/04}

\begin{document}

\title{Supporting Contraceptive Decision-Making in the Intermediated Pharmacy Setting in Kenya}

\author{Lisa Orii}
\email{lisaorii@cs.washington.edu}
\orcid{0000-0002-6490-4776}
\affiliation{%
  \institution{University of Washington}
  \city{Seattle}
  \state{Washington}
  \country{USA}
}
\author{Elizabeth K Harrington}
\email{harri@uw.edu}
\orcid{0000-0002-8196-069X}
\affiliation{%
  \institution{University of Washington}
  \city{Seattle}
  \state{Washington}
  \country{USA}
}
\author{Serah Gitome}
\email{sgitome@kemri.go.ke}
\orcid{0009-0002-3371-4213}
\affiliation{%
  \institution{Kenya Medical Research Institute}
  \city{Nairobi}
  \country{Kenya}
}
\author{Nelson Kiprotich Cheruiyot}
\email{cheruiyotnelsonk@gmail.com}
\orcid{0009-0008-5436-3102}
\affiliation{%
  \institution{Independent design consultant}
  \city{Nairobi}
  \country{Kenya}
}
\author{Elizabeth Anne Bukusi}
\email{director@kemri.go.ke}
\orcid{0000-0002-2031-2808}
\affiliation{%
  \institution{Kenya Medical Research Institute}
  \city{Nairobi}
  \country{Kenya}
}
\author{Sandy Cheng}
\email{sandyc01@cs.washington.edu}
\orcid{0009-0004-4360-539X}
\affiliation{%
  \institution{University of Washington}
  \city{Seattle}
  \state{Washington}
  \country{USA}
}
\author{Ariel Fu}
\email{arielfu@cs.washington.edu}
\orcid{0009-0003-4480-537X}
\affiliation{%
  \institution{University of Washington}
  \city{Seattle}
  \state{Washington}
  \country{USA}
}
\author{Khushi Khandelwal}
\email{khushik@cs.washington.edu}
\orcid{0009-0008-1634-1332}
\affiliation{%
  \institution{University of Washington}
  \city{Seattle}
  \state{Washington}
  \country{USA}
}
\author{Shrimayee Narasimhan}
\email{shrima.narasimhan@gmail.com}
\orcid{0009-0009-6886-0799}
\affiliation{%
  \institution{University of Washington}
  \city{Seattle}
  \state{Washington}
  \country{USA}
}
\author{Richard Anderson}
\email{anderson@cs.washington.edu}
\orcid{0000-0002-7283-7219}
\affiliation{%
  \institution{University of Washington}
  \city{Seattle}
  \state{Washington}
  \country{USA}
}

\renewcommand{\shortauthors}{Orii et al.}

\begin{abstract}
Adolescent girls and young women (AGYW) in sub-Saharan Africa face unique barriers to contraceptive access and lack AGYW-centered contraceptive decision-support resources. To empower AGYW to make informed choices and improve reproductive health outcomes, we developed a tablet-based application to provide contraceptive education and decision-making support in the pharmacy setting - a key source of contraceptive services for AGYW - in Kenya. We conducted workshops with AGYW and pharmacy providers in Kenya to gather app feedback and understand how to integrate the intervention into the pharmacy setting. Our analysis highlights how intermediated interactions - a multiuser, cooperative effort to enable technology use and information access - could inform a successful contraceptive intervention in Kenya. The potential strengths of intermediation in our setting inform implications for technological health interventions in intermediated scenarios in \lrem{LMICs}\ladd{low- and middle-income countries}, including challenges and opportunities for extending impact to different populations and integrating technology into resource-constrained healthcare settings.
\end{abstract}



\maketitle

\input{sections/01_intro}

\input{sections/02_background}
\input{sections/03_rw}
\input{sections/04_00_methods}
\input{sections/04_01_app}
\input{sections/04_02_workshop}
\input{sections/05_findings}

\input{sections/06_discussion}
\input{sections/07_conclusion}

\begin{acks}
Research reported in this publication was funded by the Eunice Kennedy Shriver Institute for Child Health and Human Development (K12HD001264) and the University of Washington Global Innovation Fund. The authors acknowledge Merceline Awuor and Cellestine Aoko for their major contributions to recruitment, data collection, and essential administrative support. We also thank Sandy Cheng, Ariel Fu, Khushi Khandelwal, Shrimayee Narasimhan, and Shobhit Srivastava for their contributions to the development of the Mara Divas app. 
\end{acks}


\bibliographystyle{ACM-Reference-Format}
\bibliography{references}


\appendix
\input{sections/08_appendix}

\end{document}

%% file: sections/01_intro.tex
\section{Introduction}

Information and communication technologies for development (ICTD) research ``involves a consideration of human and societal relations with the technological world and specifically considers the potential for positive socioeconomic change through this engagement” ~\cite{burrell2009constitutes}. ICTD research is typically set in low- and middle-income countries (LMICs) that face barriers - economic, access, literacy, and infrastructure - that impact how people interact with technologies and how technologies are developed, deployed, sustained, and received. In these settings, intermediated interactions emerge as a practice to overcome barriers to enable technology use and information access. Intermediated interaction is a multi-user cooperative effort in which digitally adept individuals (i.e., intermediaries) support individuals who are typically non-literate, lack technology-operation skills, and/or have financial constraints to use technology and access information ~\cite{10.1145/1753326.1753718, 1626204}.

We extend this model of intermediation to our work in a mobile health contraceptive project for adolescent girls and young women (AGYW) in Kenya, with experts in human-centered design, computer science/HCI, and clinical reproductive health in Kenya and the U.S. AGYW in sub-Saharan Africa face a disproportionately high risk of unintended pregnancy and poor health outcomes, and face unique barriers to contraceptive access due to their gender and age ~\cite{sully2020adding}. These challenges necessitate AGYW-centered contraceptive decision-support resources that reach this underserved and stigmatized population. Our work builds on the understanding that pharmacies are a key source of contraceptives for AGYW in Kenya ~\cite{gonsalves2023pharmacies, gonsalves2020mixed, radovich2018meets} and formative work that explores AGYW’s contraceptive preferences and needs. We developed a tablet-based application \ladd{(app) - the Mara Divas app -} that provides contraceptive education and decision-support for AGYW in the pharmacy setting in Kenya, in a way that aligns with their needs and preferences. Our app aims to not only provide recommendations for contraceptive methods that suit AGYW’s needs and preferences, but to also demystify information that is of importance to them. We created this mobile app to enable an assessment of its impact. Our motivation is to empower vulnerable AGYW to make informed contraceptive choices and contribute to improved reproductive health outcomes. 

We draw on the model of intermediation for two reasons. First, we want to understand how the substantial body of work in ICTD and human computer interaction for development (HCI4D) regarding intermediation could inform a successful intervention for AGYW in Kenya. Second, we want to apply lessons from this unique setting to inform a broader understanding of LMIC-based interventions. We address the following research questions:
\begin{itemize}
\item \textbf{RQ1:} How does the intermediation model allow us to understand how a pharmacy-based app for contraceptive education and decision-support for AGYW can be integrated into the pharmacy setting in Kenya?
\item \textbf{RQ2:} How does this experience inform approaches to the development of interventions for public intermediated scenarios?
\end{itemize}

We conducted workshops in Kisumu county, Kenya with AGYW aged 15-24 years and pharmacy staff to iteratively refine the app prototype and to understand how to integrate the intervention into the pharmacy setting. \ladd{Our findings highlight key moments in AGYW-pharmacist interactions in which pharmacists’ personal beliefs, interests, and priorities misalign with those of AGYW. We also describe differences in language, digital, and health literacy levels among AGYW subpopulations, and the critical roles of the pharmacist.} We contribute the following:

\begin{itemize}
\item \lrem{Descriptions of key moments in AGYW-pharmacist interactions, differences between subpopulations of AGYW, the app’s potential impact on AGYW-pharmacist interactions, and app feedback from participants that inspired changes.}\ladd{An articulation of the significance and role of intermediation by the pharmacist to facilitate app usage and more broadly, contraceptive education and decision-making for a wide range of AGYW.}
\item An argument that our intervention could have a differential impact on subpopulations of AGYW\ladd{, emphasizing the importance of extending impact to a hard-to-reach population}. 
\item \lrem{A reflection on implications for technological interventions in LMIC-based healthcare settings.}\ladd{Considerations for developing technologies for a complex ecology of LMIC-based healthcare, including commercial retail settings.} 
\end{itemize}

%% file: sections/02_background.tex
\section{Background}

In LMICs, nearly 50\% of adolescent pregnancies are unintended, most of which occur among adolescents who want to avoid a pregnancy but are not using a modern contraceptive method ~\cite{sully2020adding}. In East Africa, AGYW aged 15-24 years encounter multiple barriers to accessing sexual and reproductive health services, such as stigma around female sexual activity and social pressure to have a child ~\cite{sully2020adding}. Despite a need to provide contraceptive services that meet AGYW’s needs, conventional health facilities struggle to reach AGYW \ladd{because their services are expensive and time-consuming, and AGYW are concerned about healthcare providers maintaining confidentiality} ~\cite{corroon2016key}.

Commercial drug sellers, including pharmacies, are key contraceptive access points for AGYW in sub-Saharan Africa ~\cite{gonsalves2023pharmacies, gonsalves2020mixed, radovich2018meets}. Pharmacies are preferred by many Kenyan AGYW because of their convenience, privacy, speed of service, and perception of less judgemental pharmacists ~\cite{gonsalves2023pharmacies, gonsalves2020mixed}. However, pharmacy-based contraceptive services have downsides, including sporadic and inaccurate counseling ~\cite{gonsalves2020pharmacists} due to lack of pharmacy provider training ~\cite{gonsalves2019regulating} and limited availability of contraceptive methods ~\cite{gonsalves2023pharmacies}. As such, contraceptive counseling in the pharmacy setting struggles to align with AGYW’s values and preferences. The impetus of \lrem{the}\ladd{our} app is to provide person-centered contraceptive decision-support tailored to AGYW seeking contraceptive services in the pharmacy setting.

%% file: sections/03_rw.tex
\section{Related Work}

\subsection{Intermediation in LMICs}

In LMICs, where access to technology or digital literacy is limited, intermediated interactions emerge as a common practice to enable technology access and use for a vast number of people. According to Sambasivan et al. ~\cite{10.1145/1753326.1753718}, intermediation is accomplished when a digitally skilled user enables information seeking and use for a ``beneficiary-user” for whom technology is inaccessible due to non-literacy, lack of technology-operation skills, or financial constraints. Parikh and Ghosh ~\cite{1626204} articulate an understanding of intermediated tasks that identifies how technology availability, existing power relationships between users, and technical literacy can impact the ways in which interactions manifest. Prior work emphasizes the critical role of human relations or ``human infrastructure” in intermediated interactions and their potential for being ``more robust and pervasive than technology networks” ~\cite{10.1145/2369220.2369258} because they can overcome access and use constraints and can influence information penetration ~\cite{10.1145/2909609.2909664, 10.1145/1753326.1753610}. The human agent facilitating intermediation affords the flexibility to accommodate unexpected needs and behaviors ~\cite{10.1145/2737856.2738023}. However, human factors of intermediaries, such as their background and motivations, can influence intermediation ~\cite{10.1145/3449118, 10.1145/3313831.3376465, 10.1145/3411764.3445410}.

Intermediaries can support new tasks in addition to continuing their primary tasks. In Tanzania, mobile money agents, whose main purpose was to assist community members with mobile money services\lrem{, also supported} \ladd{began to support} feature phone users in using a \lrem{new }newly introduced mobile app ~\cite{10.1145/3613904.3642099}. In India, local mobile shops assisted in the dissemination of health education videos in addition to continuing their usual business ~\cite{10.1145/2909609.2909655}. Regardless of the tasks they support, intermediaries often contribute more than their technical literacy - they also facilitate information flow. In Digital Green ~\cite{4937388} and Projecting Health~\cite{10.1145/2737856.2738023} in India, mediators facilitated video-based education on domain-specific topics and discussions that furthered the audience’s understanding of the video content. In Bangladesh, small businesses, such as pharmacies and grocery shops, bridged the information gap between extremely impoverished people and low-cost healthcare services, connecting clients to affordable healthcare ~\cite{10.1145/3449118}. In these examples and beyond, intermediaries have been effective not only in supporting technology access and use, but also in disseminating information and encouraging people to engage with information ~\cite{10.1145/2369220.2369253}.

\subsection{Interventions for Person-Centered Contraceptive Education and Decision-Making Support} 

There is a growing body of work in contraceptive education and decision-support that focuses on improving preference-sensitive decision-making, meaning that method choice can depend on a variety of factors such as perception of method safety, effectiveness, and side effects ~\cite{madden2015role, marshall2016young}. Research on preference-sensitive shared decision-making with a provider during contraceptive counseling demonstrates that women and service providers differ in their preferences for and prioritization of contraceptive characteristics ~\cite{weisberg2013women}. Such work has informed the development of person-centered contraceptive decision-support tools such as My Birth Control, a web-based tool in the U.S. used before counseling to improve women’s experiences of contraceptive counseling and support them in choosing contraceptive methods that fit their needs and preferences ~\cite{dehlendorf2019cluster}. The app includes educational models and surveys to indicate preferences for method characteristics. My Birth Control increased women’s knowledge of methods, reduced decision conflict, encouraged providers to acknowledge clients’ preferences and values, and clients appeared to be more confident in their method preferences ~\cite{holt2020patient, dehlendorf2019mixed}. Our app takes inspiration from My Birth Control’s person-centered focus. 

\ladd{Person-centeredness is also evident in HCI work that explores opportunities for technology to support information seeking of sexual and reproductive health in light of the challenges and needs faced by specific populations. Dewan et al. ~\cite{10.1145/3613904.3641934} offer guidance on technology development for teenagers seeking information on reproductive health given teenagers’ information seeking practices and struggles with finding appropriate information sources. Patel et al. ~\cite{10.1145/3679318.3685380} propose technology designs that facilitate safe and appropriate sexual and reproductive care for LGBTQ+ people with uteruses, considering their experiences with in-person care.}

\lrem{However, person-centered contraceptive decision-support tools}\ladd{However, aforementioned work is situated in high-income countries. Technologies that support person-centered contraceptive decision-support} have yet to be a focus for contraceptive interventions in LMICs. There are, however, interventions such as text messaging, phone calls, or voice messaging that are designed to facilitate contraceptive education and behavior change. While these include interventions tailored to a specific population, they are intended to engage users in health-related communications, increase contraceptive knowledge, or promote contraceptive uptake and continuation, which differs from supporting person-centered decision-making ~\cite{harrington2019mhealth, mccarthy2018development, 10.1145/2702123.2702124}. A plausible explanation for the lack of person-centered contraceptive decision-making tools in LMICs may be the paucity of data on contraceptive preferences and priorities of women in these countries. \lrem{In sub-Saharan Africa, where our work takes place, t}\ladd{T}here is even less data on what adolescents consider important to their contraceptive decisions, though there are studies that demonstrate adolescents’ concerns about future fertility and side effects ~\cite{velonjara2018motherhood, ochako2015barriers, gueye2015belief}. \ladd{While there are examples of HCI-driven interventions aimed to provide adolescence related sexual and reproductive health guidance in LMICs ~\cite{10.1145/3411764.3445694, wang2022artificial}, these interventions assume that adolescents have access to personal mobile devices that can be used privately, which is not common for adolescents in many LMICs ~\cite{madonsela2023development}.}\lrem{ While there are interventions focused on contraceptive information provision}\lrem{\mbox{ ~\cite{feroz2021using}}}{\lrem{, a} \ladd{Moreover, a}pproaches to support decision-making in community-based settings, such as pharmacies where many adolescents access contraception ~\cite{gonsalves2023pharmacies, gonsalves2020mixed, radovich2018meets}, are limited ~\cite{dev2019acceptability}. Our work \ladd{aims to improve the quality and person-centeredness of contraceptive care available to AGYW by tailoring our intervention to AGYW’s needs, values, and preferences.}\lrem{ aims to reach AGYW in a setting where they access contraceptive services to improve contraceptive care.}

%% file: sections/04_00_methods.tex
\section{Methods}

This study builds on formative work aimed to provide decision-making support for AGYW seeking contraception in pharmacies in Kenya. Our team previously explored factors that influence contraceptive preferences and decision-making for AGYW in Kisumu, Kenya, assessed the relative importance of contraceptive method and service delivery characteristics, and examined the quality of contraceptive care in pharmacies in western Kenya. Rigorous qualitative and qualitative research informed our intervention design and provided an understanding of contraceptive care for AGYW in Kenya seeking contraceptive services in the pharmacy setting.

We designed and developed \lrem{an app}\ladd{the Mara Divas app} to be placed in the pharmacy setting, a key access point for contraceptive services for Kenyan AGYW, to provide contraceptive education and decision-support tailored to AGYW's needs and preferences. We conducted co-design workshops with AGYW and pharmacy staff in Kisumu county, Kenya to understand how to integrate the app into the pharmacy setting and to gather app feedback. Feedback from the workshops inspired revisions to the app, which were made in between workshops and then after. \ladd{This research was approved by the Kenya Medical Research Institute Scientific Ethics Review Unit and the University of Washington Human Subjects Division. We obtained a research permit from the Kenya National Commission for Science, Technology, and Innovation.}\lrem{This study was approved by the IRB in Kenya and at our university after undergoing rigorous review.} 

\subsection{Setting}
\ladd{This work takes place in Kisumu county which is located in western Kenya. Its largest population center is the city of Kisumu, which is the third largest city in Kenya. In 2022 in Kisumu county, 9.2\% of AGYW aged 15-19 years had ever experienced a live birth and 11.1\% had ever experienced pregnancy ~\cite{DHS2023}. Luo is the primary ethnic group of Kisumu county where DhoLuo, Kiswahili (the national language), and English are spoken. Public health clinics offer most contraceptive methods for free or for a small fee. Pharmacies usually offer condoms, emergency contraceptives, depot medroxyprogesterone acetate (injectables, also known as ``depo”), and combined oral contraceptive pills. Some pharmacies also place contraceptive implants on site, but most refer patients to nearby clinics for implants or intrauterine devices (IUD/IUCD).}

%% file: sections/04_01_app.tex
\subsection{App Design and Development}
Considering that AGYW have low levels of knowledge regarding side effects of contraception and hold strong beliefs about contraceptive harms ~\cite{harrington2021spoiled, ochako2015barriers, sedlander2018they}, we did not develop an app that only recommended the ``right” contraceptive method. We were motivated to provide information on various contraceptive options regarding topics that are important to AGYW, including bleeding patterns, privacy of contraceptive use, and influence on future fertility ~\cite{harrington2023adaptation, harrington2021spoiled} and demystify side effects. We were inspired by the person-centeredness of My Birth Control, a U.S.-based decision support tool that assists women with contraceptive method selection ~\cite{dehlendorf2019cluster, holt2020patient, dehlendorf2019mixed}. \ladd{We named our app the Mara Divas app, in which ``Mara” means ``my own” in DhoLuo, the local language of Kisumu county, Kenya.}

The \ladd{Mara Divas} app is \lrem{intended}\ladd{designed} to be used in a private counseling room in the pharmacy setting. \ladd{The room will provide a tablet with the app installed and a pair of headphones. The intended users of the app are AGYW seeking contraceptive services in the pharmacy setting. As such, when we use the term ``user” henceforth, we refer to the AGYW using the app.} When the AGYW approaches the pharmacist, the pharmacist would take them to a private counseling room to use the app for 5-20 minutes. Lastly, the pharmacist would communicate directly with the AGYW about their choice and provide either the desired method or a referral.

\begin{figure*}[hbt!]
  \centering
\includegraphics[width=1.0\linewidth]{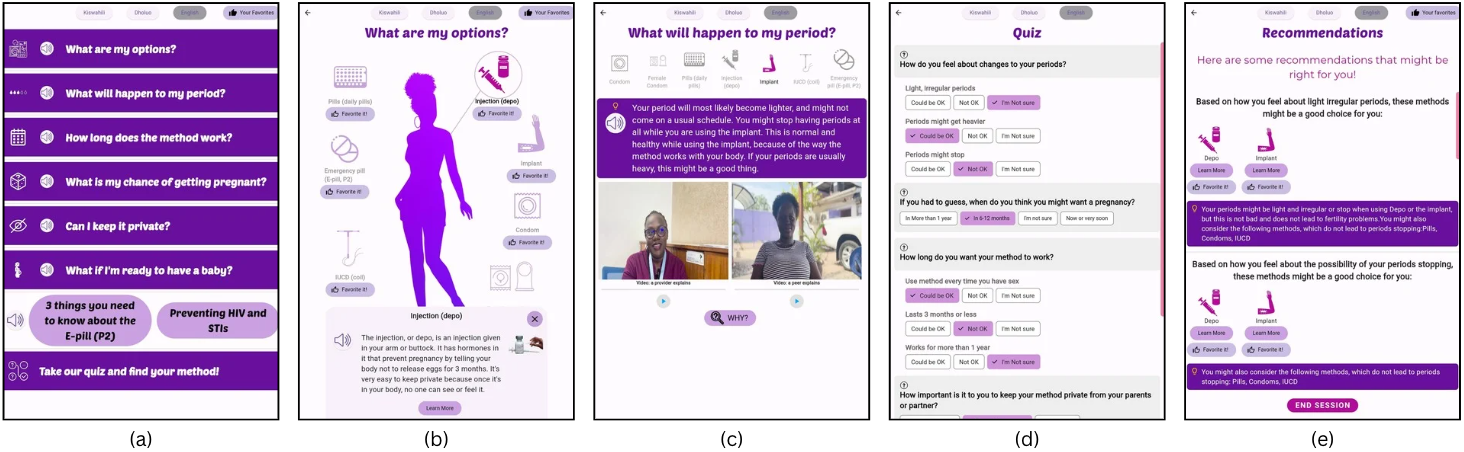}
\caption{Mara Divas app configuration. Enlarged images of each screen are in Appendix \ref{big_images}. (a) Main screen with questions and topics that are of most interest and importance to AGYW. (b) Short explanations of each contraceptive method and where it is administered on a female body. (c) Information presented with text, audio, and videos (provider and peer). (d) Survey where users can indicate preferences for method characteristics. (e) Recommendations for contraceptive methods based on survey results.}
\Description{Five app screens, described from left to right. (a) Main screen with nine questions and topics that are of most interest and importance to AGYW. Top of the screen has three language toggle buttons labeled ``Kiswahili,” ``Dholuo,” and ``English” where ``English” is grayed out to indicate its selection. These buttons are presented in all of the screens explained after. Top of the screen also includes a ``Your Favorites” button with a thumbs-up sign preceding the text, which is also in screens (b) and (e). The screen is filled with buttons, labeled in order from top to button, ``What are my options?” ``What will happen to my period?” ``How long does the method work?” `1`What is my chance of getting pregnant?” ``Can I keep it private?” ``What if I’m ready to have a baby?” and ``Take our quiz and find your method!” Each button is accompanied by a representative image of that topic and an audio button. Two buttons labeled ``3 things you need to know about the E-pill (P\ladd{-}2)” and ``Preventing HIV and STIs” is also exhibited. (b) Top of the page is titled ``What are my options?” A female body in the center of the screen with contraceptive method icons around it. On the left side of the body are icons and accompanying labels of pills (daily pills), emergency pill (e-pill, P\ladd{-}2), IUCD (coil), and on the right side are injection (depo), implant, condom, and female condom. Below each label is a button ``Favorite it!” with a thumbs-up sign preceding the text. The injection (depo) is selected, indicated by a different color. A pop-up at the bottom of the screen shows a description of the injection (depo) with an audio button and a real image of the injection. (c) Top of the page is titled ``What will happen to my period?” Below the title are seven contraceptive methods with icons and labels, as described in screen (b). Implant is selected, indicated by a different color. Middle of the screen displays a text description of the method and an audio button. Below this is two videos - one of a provider, one of an AGYW peer - placed side by side. Below the videos is a button labeled ``WHY?” with a magnifying glass icon preceding the text (d) Top of the page is titled ``Quiz.” Multiple choice quiz/survey questions displayed are ``How do you feel about changes to your periods?” ``If you had to guess, when do you think you might want a pregnancy?” ``How long do you want your method to work?” and ``How important is it to you to keep your method private from your parents or partner?” Selections for each question are indicated by a different color. A scroll bar indicates continuation of the quiz. (e) Top of the page is titled ``Recommendations” and below it, ``Here are some recommendations that might be right for you!” Recommendations displayed are titled ``Based on how you feel about light, irregular periods, these methods might be a good choice for you:” and ``Based on how you feel about the possibility of your periods stopping, these methods might be a good choice for you:” with icons and labels of the depo and implant. Below each method label is a button labeled ``Learn More” and ``Favorite it!” with a thumbs-up sign preceding the text. Below the buttons are short relevant explanations for other recommendations. The bottom of the screen has a button labeled ``END SESSION.” A scroll bar indicates continuation of the quiz.}
\label{fig:app}
\end{figure*}

We built an Android app using the Flutter \ladd{3.19.5} platform\ladd{~\cite{flutter} that supports Android 5.0 and all subsequent versions. Our app targets} \lrem{targeting }Samsung Galaxy \ladd{Tab A8} tablets \ladd{with a 10.5-inch display}. We chose tablets because the pharmacy setting provides a large enough space for private table usage and the contents on a tablet would be easier to interact with than on a phone. We chose Android devices because of their low cost and AGYW’s familiarity with Android. Flutter was selected as an app development framework because it is a mature mobile app platform and provides a professional app appearance.


The content of the app was carefully crafted by a clinical reproductive health expert who has extensive research experience in Kenya. The app’s main screen presents sections with topics that are of most interest to AGYW (Figure \ref{fig:app} a), as learned through formative work. Each section leads the user to a screen with topic-specific information for seven contraceptive methods (Figure \ref{fig:app} b, c): male condom, female condom, daily contraceptive pills, implant, injection (depo), IUCD, and emergency contraceptive (e-pill). For each method, we also made a screen that briefly summarized information about that method. Users can also ``favorite” methods that they like and want to return to for review. 

Once all sections are visited, users complete a survey to indicate preferences for method characteristics (Figure \ref{fig:app} d). The survey is composed of five multiple choice questions. Although five questions are limited in understanding the users’ contraceptive preferences, we restricted the number of questions because AGYW dislike lengthy questioning ~\cite{gonsalves2020pharmacists}. Based on survey responses, the app recommends contraceptive methods by the user’s method characteristic preference (Figure \ref{fig:app} e). 

We prepared the \ladd{Mara Divas} app in DhoLuo, Kiswahili (the national language), and English, which are languages spoken in Kisumu county, Kenya, where the workshops took place and the app is planned to be piloted. A language toggle tool is available on all screens so that users can engage with content in their preferred language. To reach AGYW with limited language literacy, we included audio and brief video components (Figure \ref{fig:app} c). Text is accompanied by a spoken version of the text which was recorded by study staff, indicated by an audio button next to the text. We included two types of videos: videos where providers share their expertise on contraceptive topics, and videos where AGYW share their experiences with using contraceptives. These videos were narrated by service providers and AGYW actors in Kisumu. We incorporated four provider videos and three peer videos. Videos were between one to three minutes long.

%% file: sections/04_02_workshop.tex
\subsection{Co-design Workshops}
We conducted co-design workshops to understand how to integrate the intervention into the pharmacy setting and to gather potential app \lrem{users}\ladd{stakeholders, namely AGYW and pharmacy staff,} with varied preferences to share feedback on the app and ideate new designs or content. 

\subsubsection{Study Design}
\lrem{The workshops took place in Kisumu county, Kenya. Luo is the primary ethnic group of Kisumu county where DhoLuo, Kiswahili, and English are spoken. Public health clinics offer most contraceptive methods for free or for a small fee. Pharmacies usually offer condoms, emergency contraceptives, depot medroxyprogesterone acetate (injectables, also known as ``depo”), and combined oral contraceptive pills. Some pharmacies also place contraceptive implants on site, but most refer patients to nearby clinics for implants or intrauterine devices (IUD/IUCD).} 

We conducted four workshops - three with AGYW aged 15-24 years with pharmacy-based contraceptive care experience and one with pharmacy staff - to allow for iteration in app prototype development between workshops. Workshops were conducted between April - May 2024. Workshops were led by four facilitators, all who spoke English and Kiswahili and two who also spoke DhoLuo. Facilitators were also involved in the creation and revision of the study design. Workshop activities were mainly led by two facilitators with human-centered design experience (third and fourth authors). The first author was also present for the first three workshops. 

Each workshop began with a distribution of informed consent forms, followed by an explanation of consent guidelines by study staff and a pre-workshop questionnaire administered on a tablet by study staff. After the pre-workshop questionnaire, participants re-grouped to review the background and objectives of the study, and confidentiality and consent guidelines. We then proceeded to conduct activities where participants engaged in discussion and produced artifacts (e.g., post-it notes) and facilitators took field notes. Workshops lasted between 9AM and 5PM, including \lrem{a }\ladd{one 90-minute} lunch break and \ladd{two 30-minute} tea breaks \ladd{in between workshop activities}. Numerous small revisions to the app were made in response to feedback in between workshops and after completion of all workshops. 

\subsubsection{Participant Recruitment}
AGYW participants were recruited using community-based strategies to optimize the diversity of perspectives and experiences in the three AGYW workshops by age, education, contraceptive experience and experience with mobile apps. Purposive sampling ~\cite{tongco2007purposive} was used to recruit AGYW from peer mobilizers in the community and in-person in the pharmacy settings. Snowball sampling ~\cite{parker2019snowball} was also used, by which study staff asked interested participants to recommend peers for inclusion. Female study staff who spoke DhoLuo, Kiswahili, and English provided an introduction to the study and potential participants who met the inclusion criteria were assessed for their availability for workshop dates. AGYW under 18 years of age and not emancipated minors were encouraged to speak to a trusted adult about participating in the study. Eligible AGYW met the following criteria: 1) cisgender female; 2) age 15-24 years; 3) has accessed pharmacy-based contraception in the last three months; 4) not currently pregnant; 5) speaks DhoLuo, Kiswahili, and/or English; 6) feels comfortable reading DhoLuo, Kiswahili, and/or English; and 7) willing to sign informed consent. \ladd{This work targets AGYW aged 15-24 years because we determined that this age group is representative of a population that will most likely benefit from our app. We also note that most AGYW under the age of 15 are not sexually active, as only 8\% of Kenyan AGYW aged 15-24 years experienced first sexual intercourse before the age of 15 ~\cite{kenyadhs}. We also wanted to minimize the risks to a vulnerable population by working with slightly older AGYW.}

Pharmacy staff participants were recruited from pharmacies in Kisumu and periurban areas to balance the gender and age of pharmacy staff in the pharmacy staff workshop. Eligible pharmacists met the following criteria: 1) at least 18 years of age; 2) currently working in a pharmacy providing condoms, emergency contraception, combined oral contraceptive pills, and depot medroxyprogesterone acetate (hormonal contraceptive injection); 3) currently working in a pharmacy that has a counseling room/private space; 4) speaks Kiswahili and/or English; 5) feels comfortable reading DhoLuo, Kiswahili, and/or English; and 6) willing to sign informed consent. 

We recruited 33 AGYW and 10 pharmacy staff (6 female). \ladd{We conducted four workshops:}\lrem{We had} one workshop of 12 AGYW aged 15-17 years\footnote{\lrem{There was one 19-year-old in the workshop with a younger age group but}\ladd{All except one participant in the younger age group workshop was aged 15-17 years and so} we present the \lrem{two }\ladd{AGYW} workshops as having \ladd{two} distinct age groups.}\ladd{,} \lrem{and} two workshops with 10 and 11 AGYW aged 18-24 years\ladd{, and one workshop of 10 pharmacy staff}. The sample size was determined through the review of relevant literature in the health domain using similar methods ~\cite{hunter2021designing, wilkinson2022developing, harrington2021spoiled} and guidelines in qualitative research ~\cite{sandelowski1995sample}. AGYW ranged in age from 15-24 years \lrem{(mean=19.24, SD=2.75)}\ladd{and the median age was 19 (IQR 17-21).}\lrem{ and p} \ladd{P}harmacy staff ranged in age from 32-48 years \lrem{(mean=37.10, SD=6.87)}\ladd{and the median age was 33.5 (IQR 32-40)}. \ladd{A summary of participant characteristics for the three AGYW workshops and one pharmacy staff workshop are in Tables \ref{tab:AGYW-participants} and \ref{tab:pharm-participants}, respectively.} AGYW participants and \lrem{pharmacist}\ladd{pharmacy staff} participants were compensated with 1000 KSH and 2000 KSH (approximately 8 USD and 15 USD), respectively, for their participation. \ladd{The compensation was determined based on the judgment of the Kenyan ethics review board of what was appropriate and not coercive.}

\begin{table}[h!]
\caption{Participant characteristics of AGYW workshops.}
\Description{Table with characteristics and workshop ID for 3 AGYW workshops. Characteristics are in one column with median age (IQR), education level, and mobile phone use. Corresponding values are in the other columns where each column represents one workshop.}
\label{tab:AGYW-participants}
\resizebox{\columnwidth}{!} {
\begin{tabular}{|p{4.4cm}||p{1.6cm}|p{1.6cm}|p{1.6cm}|}
\hline
\multicolumn{1}{|c||}{} &
\multicolumn{3}{c|}{\textbf{Workshop ID}} \\
\multicolumn{1}{|c||}{\textbf{Characteristic}} &
\multicolumn{1}{c}{\textbf{AGYW-1}} &
\multicolumn{1}{c}{\textbf{AGYW-2}} &
\multicolumn{1}{c|}{\textbf{AGYW-3}} \\
\hline
Total Participants, n & 10 & 12 & 11 \\
\hline
Median Age (IQR) & 20.5 (19.25-21.75) & 17 (15.75-17) & 21 (19-23) \\
\hline
Education Level, n &  &  & \\ 
\hspace{3mm}Primary school, not complete & 0 & 1 & 0 \\ 
\hspace{3mm}Primary school, complete & 1 & 3 & 1 \\
\hspace{3mm}Secondary school, not complete & 1 & 7 & 1\\
\hspace{3mm}Secondary school, complete & 3 & 1 & 7\\
\hspace{3mm}Post-secondary school & 5 & 0 & 2 \\ 
\hline
Mobile Phone Use, n &  &  & \\ 
\hspace{3mm}Has access to a mobile phone & 10 & 5 & 11 \\ 
\hspace{3mm}Owns a personal mobile phone & 10 & 4 & 11 \\
\hspace{3mm}Has access to a smartphone & 9 & 2 & 6 \\
\hline
\end{tabular}
}
\end{table}

\begin{table}[h!]
\caption{Participant characteristics of pharmacy staff workshop.}
\Description{Table with characteristics and workshop ID for pharmacy staff workshop. Characteristics are in one column with median age (IQR), education level, and years of professional experience. Corresponding values are in the other columns where each column represents one workshop.}
\label{tab:pharm-participants}
\resizebox{\columnwidth}{!} {
\begin{tabular}{|p{7cm}||p{4cm}|}
\hline
\multicolumn{1}{|c||}{} &
\multicolumn{1}{c|}{\textbf{Workshop ID}} \\
\multicolumn{1}{|c||}{\textbf{Characteristic}} & \multicolumn{1}{c|}{\textbf{Pharmacy-1}} \\
\hline
Total Participants, n& 10 (6 female) \\
\hline
Median Age (IQR) & 33.5 (32-40) \\
\hline
Licensure, n & \\
\hspace{3mm} Pharmacy owner & 1\\
\hspace{3mm} Pharmacy technician & 8\\
\hspace{3mm} Other & 1\\
\hline
Years of Working in the Pharmacy, n &   \\ 
\hspace{3mm}5 or more years & 9 \\ 
\hspace{3mm}1-2 years & 1 \\
\hline
Median Score for Confidence in Advising AGYW (IQR) &   \\ 
\hspace{3mm}On family planning options & 5 (4-5) \\ 
\hspace{3mm}On how to use family planning methods & 5 (4-5) \\
\hspace{3mm}On family planning side effects & 5 (4.25-5) \\
\hline
\end{tabular}
}
\end{table}

\subsubsection{Data Collection}
\begin{figure*}[hbt!]
  \centering
\includegraphics[width=1.0\linewidth]{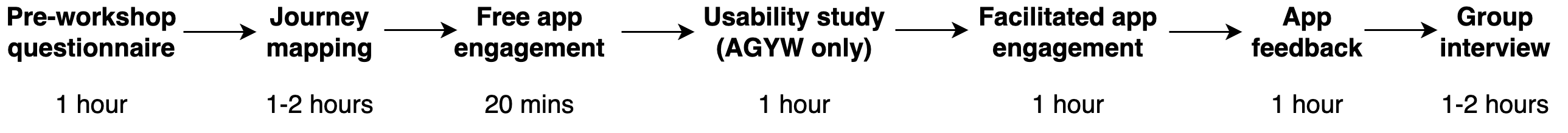}
\caption{Flow of workshop activities.}
\Description{Flow of workshop activities from left to right: pre-workshop questionnaire (1 hour), journey mapping (1-2 hours), free app engagement (20 mins), usability testing (1 hour), facilitated app engagement (1 hour), app feedback (1 hour), group interview (1-2 hours).}
\label{fig:flow}
\end{figure*}

Data was collected through pre-workshop questionnaires, artifacts (e.g., post-it notes), \ladd{written} field notes \ladd{in English, and audio-recordings. The primary data sources were artifacts and written notes that were produced for each workshop and all its activities, as the workshops were centered around small or large group activities that involved creating visual representations of ideas and concepts.}\lrem{, and audio-recordings of} \ladd{Large group interviews were audio-recorded as a secondary source of data to retain individual participant quotations and perspectives.} \lrem{ facilitated app engagement, app feedback, and group interviews.}\ladd{When participants were engaged in small-group activities or speaking one-on-one with a facilitator, it was impractical to rely on audio-recordings, so written note-taking was used.} Activities were conducted in the order they are presented in this section. \ladd{The flow of activities is described in Figure \ref{fig:flow}.}

\paragraph{Pre-workshop questionnaire}
Each participant individually completed a pre-workshop questionnaire with study staff. For AGYW, we collected 1) demographic information; 2) mobile phone and app use; 3) relationships and sexual history; and 4) reproductive history and contraceptive experience and desires. In this paper, we present data on aspects of AGYW’s demographic \lrem{information}\ladd{characteristics including age and education levels}, mobile phone \lrem{and app use}\ladd{use}, and contraceptive experiences. Other data will be presented in a future paper. For pharmacy staff, we collected 1) demographic information; and 2) confidence in advising AGYW on contraception. \ladd{For confidence in advising AGYW on contraception, we used a 5-point Likert scale, with 1 being not at all confident and 5 being extremely confident.} The pre-workshop questionnaire is included in Appendix \ref{question}. 

\paragraph{Journey mapping}
Facilitators led participants to create a journey map to visualize the entire flow of a pharmacy visit experience. We focused on ``pain points” - moments that create discomfort, embarrassment, or inconvenience - and moments where there is the potential for delight. Journey mapping lasted between one to two hours. 

\begin{figure*}[hbt!]
  \centering
\includegraphics[width=1.0\linewidth]{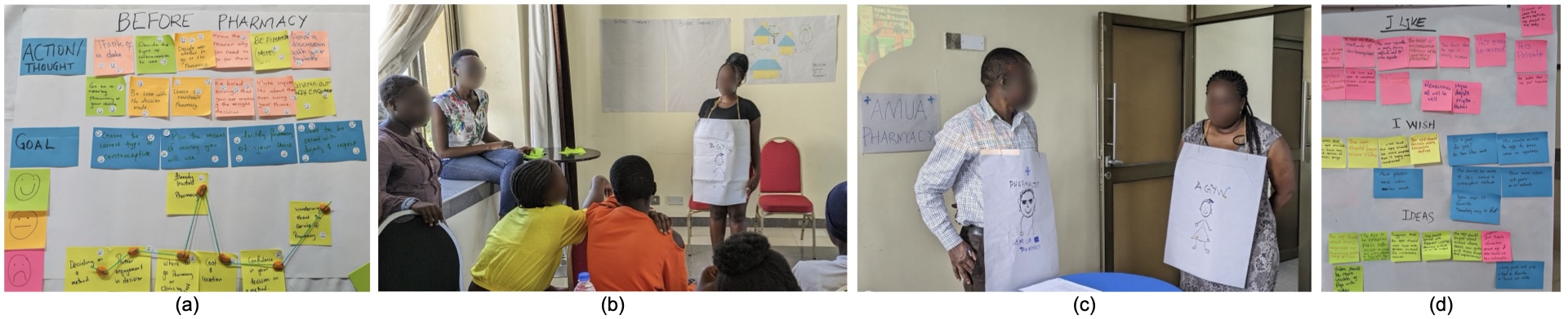}
\caption{Workshop activities. Enlarged images are in Appendix \ref{big_images}. (a) AGYW journey mapping where AGYW and facilitators wrote actions, thoughts, goals, and emotions before a pharmacy visit on post-it notes. (b) AGYW journey mapping with role-play, led by the facilitator. (c) Pharmacy staff journey mapping with role-play. (d) App feedback documented by AGYW on post-it notes using the ``I Like, I Wish, What If” method.}
\Description{Four images of workshop activities, described from left to right. (a) Poster paper with "BEFORE PHARMACY" at the top. Below, a post-it note labeled "ACTION/THOUGHT" and next to it multiple post-it notes with actions and thoughts AGYW have before visiting the pharmacy setting. Below is a post-it note labeled "GOAL" and next to it post-it notes of goals that AGYW have for visiting the pharmacy setting. Below are three post-it notes placed vertically, each with a different facial expression drawn on it, where one is happy, one is neutral, and one is sad. Post-it notes are placed next to it and connected with a string to indicate the emotional path. On each post-it note is a sticker with the same three facial expressions. (b) A girl wearing a sign labeled "AGYW" with a drawing of a girl is facing five girls (workshop participants) and a woman (workshop facilitator). (c) A sign on the wall that says "AMUA PHARMACY". A man wearing a sign labeled "Pharmacist Amua Pharmacy" with a drawing of a man is talking to a woman with a sign labeled "AGYW" and a drawing of a girl. A paper labeled ``AMUA pharmacy” is behind them, to illustrate that the characters are in a pharmacy. (d) Poster paper with "I LIKE" at the top with post-it notes of comments on what participants liked about the app, "I WISH" with post-it notes with comments on  what participants wished the app had, and "IDEAS" with post-it notes with comments on ideas participants had with improving the app.}
\label{fig:workshop}
\end{figure*}

For the AGYW workshop\ladd{s}, we applied different methods of journey mapping. In some workshops, participants used post-it notes to write their actions, thoughts, and goals that they have before, during, and after pharmacy visits (Figure \ref{fig:workshop} a). In other workshops, participants conducted an interactive role-play depicting a pharmacy visit, in which one participant acted as the pharmacist, another acted as an AGYW, and observing participants commented on their experiences during the role play (Figure \ref{fig:workshop} b). The purpose was to understand counseling from AGYW’s perspective.

In the pharmacy staff workshop, one participant acted as a pharmacist and one member of the study team acted as an AGYW to role-play a contraceptive counseling scenario (Figure \ref{fig:workshop} c). The purpose was to understand how a pharmacy staff would conduct counseling. After role-play, participants discussed how their experiences differed or aligned with what was presented.

\paragraph{Free app engagement}
Facilitators introduced the \ladd{Mara Divas} app to the participants, explaining its objective and the workshop’s goal to gather app feedback. Each participant was given a tablet and headphones. Participants were given unstructured time to engage with the app to gain familiarity. Workshop facilitators answered questions and documented real-time observations and feedback. Free app engagement \ladd{was conducted in both AGYW workshops and the pharmacy staff workshop. This activity} lasted approximately 20 minutes.

\paragraph{Usability testing}
In AGYW workshops, participants individually conducted a series of tasks on the app. The purpose was to test how easy the app was used by AGYW to inform changes for future design iterations. Each participant was given three tasks to complete (e.g., watch a video in DhoLuo of a peer talking about how to keep family planning methods private) and a facilitator observed and took notes as the participant attempted the tasks. We prepared 12 tasks total. \ladd{Usability testing was only conducted in the AGYW workshop}\lrem{We did not conduct usability testing in the pharmacy staff workshop} because pharmacy staff were not target users \ladd{of the app}. Usability testing lasted approximately one hour.

\paragraph{Facilitated app engagement}
As a group, participants discussed specific aspects or sections of the app. Participants were asked to discuss whether and how their opinions on contraception changed after engaging with the app, whether the content was helpful for addressing AGYW concerns, and how the app could influence their decision-making. In the pharmacy staff workshop, participants \ladd{also} discussed whether and how the app’s content differed from their training or experience. All participants also discussed the language toggle tool, audio, and video and compared their experiences and preferences with reading, listening, and watching. Discussions were audio-recorded. \ladd{Facilitated app engagement was conducted in both AGYW workshops and the pharmacy staff workshop.} This activity lasted approximately one hour.

\paragraph{App Feedback}
In a group, facilitators elicited structured positive feedback, criticisms, questions, and ideas regarding the app. We used the “I Like, I Wish, What If” method ~\cite{Dam_Siang_2023} because we thought this would be a helpful way to engage in participatory ideation for participants who do not have much experience with giving constructive critique. 
Participants and/or facilitators recorded the feedback on post-it notes (Figure \ref{fig:workshop} d). Discussions were audio-recorded. \ladd{App feedback was conducted in both AGYW workshops and the pharmacy staff workshop.} This activity lasted approximately one hour.

\paragraph{Group interview}
Facilitators conducted a semi-structured group interview to discuss how participants’ experiences in the pharmacy would change after exploring the app. Participants discussed their concerns with app usage in the pharmacy setting, the pharmacist’s role in supporting AGYW with and after using the app, how to combine counseling with app engagement, the app’s usefulness to the pharmacy staff, and how contraceptive access can be made easier for AGYW. Discussions were audio-recorded. \ladd{The group interview was conducted in both AGYW workshops and the pharmacy staff workshop.} This activity lasted approximately one to two hours. 

\subsection{Data Analysis}

Data sources included detailed written notes \ladd{in English} recorded by a designated staff member during the workshops, selective transcriptions of workshop audio-recordings, and workshop artifacts (e.g., arranged post-it notes from journey maps and app feedback, photos). \ladd{The primary data sources were artifacts and written notes that were produced for all workshop activities and the secondary data source was audio-recordings from large group workshop activities.} 

Multimedia data from the workshops were analyzed using two processes in tandem: a collaborative analysis by the larger design team (including the first, second, third, and fourth authors), and an in-depth thematic analysis by the first author. Team-based collaborative analysis took place in real time immediately after each workshop and at weekly design team meetings \ladd{that} continued after workshop completion. The collaborative team was made up of Kenyan study staff and researchers from Kenyan and U.S. institutions, with expertise in human-centered design, computer science\ladd{/HCI}, and clinical reproductive health. In the meetings immediately after and between workshops, the team discussed, compared, and contrasted observations within and among workshops. The main goal of these discussions was to identify app design updates to be made for the next workshop, using key insights generated from workshop activities such as free app engagement, usability testing, facilitated app engagement, and app feedback. After data collection was complete, the collaborative design team conducted analytic virtual meetings to consolidate data sources and begin to interpret the data holistically. The team recollected and discussed workshop interactions, observations, and feedback from participants that was particularly insightful, emerged frequently, or was surprising, and cataloged the iterative changes to the app. The Kenyan team members contributed additional expertise in the local context and languages, which enriched and contextualized the data.

Concurrent with the collaborative process, the first author qualitatively analyzed the \lrem{data}\ladd{artifacts and the written notes} with a specific focus on AGYW interactions with the tablet-based app and perceptions of the role of the pharmacist with the app. Using an inductive approach, the first author independently reviewed the \lrem{data}\ladd{the artifacts and the written notes} from the journey mapping, with a focus on key actions and emotions that AGYW and pharmacists expressed, and created an initial matrix of concepts highlighting AGYW concerns, desires, and decision-making processes when accessing contraception in the pharmacy setting. These concepts contextualized participant responses generated from app-specific workshop activities\lrem{Transcripts from the workshops, alongside text from w} (i.e., free and facilitated app engagement, app feedback, group interview), which were coded using an inductive approach guided by the intermediation model. Using thematic analysis ~\cite{Braun_Clarke_2022}, the codes were consolidated into an initial set of themes and iteratively refined. Themes were then discussed between the first and last author to build consensus and develop cross-cutting insights across the workshops. \ladd{After identifying themes, the first author reviewed the audible audio-recordings to select representative participant quotations that speak to the themes and concepts and transcribed and translated them in English with support from the Kenyan study staff.}

\subsubsection{Ethical Considerations}
Although minors were involved in this study, parental permission for study involvement was waived with approval from the \lrem{IRBs}\ladd{ethics review boards} in Kenya and the U.S. This decision was made through reflection on the reality that parental involvement in sensitive sexual and reproductive health research among adolescents may be harmful for adolescent research participants. To prepare to address conflicts arising from parents who have concerns about their child’s participation in the study, we encouraged minors to consult with a trusted adult prior to giving their assent. Participants were given a contact number so that parents may call to learn more or to arrange to talk with the study team. 

%% file: sections/05_findings.tex
\section{Findings}

We first present a summary of participant demographics. \lrem{Among 21 AGYW aged 18-24 years, seven were in post-secondary school, 10 had secondary school as their highest level of education, two entered secondary school but have not/did not complete, and two had primary school as their highest level of education. Among 12 AGYW aged 15-17 years, eight entered secondary school but have not/did not complete and four had primary school as their highest level of education. }In this paper, we refer to ``older AGYW” as those aged 18-24 years \ladd{(i.e., workshops AGYW-1 and AGYW-3) }and ``younger AGYW” as those aged 15-17 years \ladd{(i.e., workshop AGYW-2). Table \ref{tab:AGYW-participants} details participant characteristics for the three AGYW workshops. Younger AGYW, as expected, had overall lower levels of education than older AGYW. While all older AGYW owned a personal mobile phone, less than half of younger AGYW owned a personal mobile phone. Smartphone access was more common among older AGYW. This information regarding AGYW’s ownership of and access to a smartphone validates our decision to place it in the pharmacy setting, which we discuss in detail in section \ref{changes-not-made}}. Among all 33 AGYW, the most common contraceptive method that was used was condoms (n=25), followed by emergency contraceptive (n=13) and implants (n=13). 
\lrem{Among 22 AGYW aged 18-24 years, all owned or shared a mobile device with family and 16 used a smartphone. Among AGYW who owned or shared a mobile device, the median score for the level of comfort using mobile apps on a 5-point Likert scale was 4. On the other hand, among 12 AGYW aged 15-17 years, only five owned or shared a mobile device and only two used a smartphone.} 

\ladd{As indicated in Table \ref{tab:pharm-participants}, the 10 pharmacy staff’s confidence levels in advising AGYW were high. We measured the pharmacy staff’s confidence levels on advising AGYW on family planning options, on how to use family planning options, and on family planning side effects on a 5-point Likert scale, with 1 being not at all confident and 5 being extremely confident. The median scores were 5 (IQR 4-5), 5 (IQR 4-5), and 5 (IQR 4.25-5), respectively.} \lrem{Among 10 pharmacy staff, nine have been working in the pharmacy setting for five or more years and one has been working for one to two years. The median scores for confidence in advising AGYW on contraceptive options, advising AGYW on how to use contraceptive methods, and advising AGYW on contraceptive side effects were all 5. }

\lrem{In this section}\ladd{Next}, we first provide a summary of app feedback and how they impacted intervention decisions. Then, we elaborate on findings from group discussions and thematic analysis that expanded our understanding of how the \ladd{Mara Divas} app could integrate into the pharmacy setting. \ladd{In our findings, we refer to the workshop IDs except when they are obvious in the context of the findings (i.e., findings from younger AGYW are from workshop AGYW-2, findings from pharmacy staff are from workshop Pharmacy-1.)}

\subsection{App Feedback}

\subsubsection{Positive validation on app design}

Overall, we received positive app feedback. Pharmacists said that the app answered AGYW’s commonly asked questions in a \textit{``systematic”} way that helped \lrem{users step}\ladd{with stepping} through questions. AGYW also said that the app covered topics of importance and interest to them. AGYW and pharmacists appreciated the videos featuring providers, saying they added credibility to the app’s information. AGYW also praised the peer videos, noting how they felt encouraged when hearing their peers talk confidently about their contraceptive experiences. \ladd{AGYW also appreciated the provision of headphones, saying that they promoted privacy so that no one in the pharmacy could know what the AGYW is listening to on the app (AGYW-3).}

\subsubsection{Changes made}

Aside from significantly changing the app’s color scheme to bright purple and pink based on feedback from AGYW, most changes were to do with content or information presentation. One notable change is in the framing of the app. Pharmacists and AGYW said that ``family planning” was not relatable for AGYW who were not preparing to raise a family, but rather were prioritizing pregnancy prevention\lrem{, as summarized by a pharmacy staff}: \textit{``What is family planning, by the way? You are planning a family. This is a 16-year-old person who does not have a family. What are they planning? `I want to prevent pregnancy\lrem{.}’”} \ladd{(Pharmacy-1).} Prior work also reflects AGYW’s perceived dissonance between ``family planning” and ``pregnancy prevention” ~\cite{harrington2021spoiled}. To make the app’s framing more appropriate for AGYW’s needs and desires, we replaced the phrase ``family planning” with ``pregnancy prevention” throughout the app. 

We made signifiers on the app more findable and recognizable (e.g., larger buttons, text that prompts actions). Younger AGYW had difficulty understanding the icons that represented contraceptive methods and did not find contraceptive method labels (e.g., “IUCD (coil)”) helpful, likely because of their low language literacy. We complemented the icons with real-life images of methods so that AGYW of varying literacy levels can comprehend app content.

We also made changes to information presentation techniques. AGYW had difficulty finding information regarding how to prevent HIV and STIs and important things to know about the emergency contraceptive. Although this information was linked from other screens, many AGYW were unable to navigate through multiple screens to find this information. We added a button on the main screen (Figure \ref{fig:app} a), where all main topics were outlined, that linked directly to the information. This change was also made in light of pharmacists’ emphasis on the importance of educating AGYW about preventing HIV and STIs.

\subsubsection{Changes not made}
\label{changes-not-made}

In all workshops, participants suggested making the \ladd{Mara Divas} app available on the app store so that more people can access it. However, our intention is to restrict the app’s availability to the pharmacy setting. The main reason was AGYW’s lack of smartphone access, \lrem
{which was confirmed with}\ladd{as evident in} younger AGYW \ladd{participant demographics} \lrem{in our workshops }\ladd{(Table \ref{tab:AGYW-participants})}. We also considered the lack of privacy when using this app on personal or shared mobile devices\ladd{. Some AGYW have access to a mobile device but share it with family members. If the app were made available on the app store and installed on shared mobile devices, there is a risk that other users would discover that the AGYW is sexually active, which could cause harm to them. We were motivated to make our app low-barrier by placing it in}\lrem{ and the benefits of using the app in} a pharmacy’s private counseling room where AGYW would not be seen. \ladd{Another reason for restricting this app to the pharmacy setting is that at this research stage, the app is too premature to make it publicly available. The app and its effectiveness needs to be studied and refined to make it usable, understandable, and beneficial for AGYW before making it widely available.}

Participants from all of the workshops expressed that the app should include information on male contraceptives and/or aim to educate male peers. Aside from the lack of time to incorporate these ideas into the final version of the app, we did not add information targeting male youth because such feedback fell outside of the scope of the motivation of this app, namely focusing on supporting AGYW (not young boys or men) in contraceptive education and decision-making. 

Older AGYW also suggested adding all languages spoken in Kenya to the app. Aside from the lack of time to translate all content in all languages, we restricted the app to three languages because most AGYW in Kisumu county would only be comfortable with DhoLuo, Kiswahili, and/or English.

\subsection{Key moments in counseling} 

We shed light on pharmacist behaviors and attitudes and the realities of AGYW-pharmacist interactions that give context to challenges in pharmacy-based counseling. 

\subsubsection{Pharmacists’ personal beliefs affect AGYW-pharmacist interactions} 

We highlight key moments of the role play in the pharmacy staff workshop \ladd{(Figure \ref{fig:workshop} c)} to present examples of misalignment between pharmacists’ personal beliefs, interests, and priorities and those of AGYW. The entire role play lasted approximately 20 minutes.

\begin{enumerate}
\item Pharmacist asks AGYW whether her religion allows her to use contraceptives.
\item Pharmacist asks AGYW about her boyfriend, insists that she brings her boyfriend to her next visit, and asks for her boyfriend’s contact information. AGYW says she does not want her boyfriend involved, but the pharmacist insists.
\item Pharmacist is concerned about the AGYW’s potential exposure to HIV and STIs and risk of dropping out of school. AGYW says she is not concerned about HIV and STIs, only about getting pregnant.
\end{enumerate}

\ladd{In each of these key moments, there is misalignment between what the AGYW wants and what the pharmacist insists. The pharmacist’s priorities for the AGYW are also apparent, such as staying in school and minimizing exposure to HIV and STIs. The incongruence between the priorities of the AGYW and those of the pharmacists is one of the reasons why AGYW are anxious about approaching the pharmacy setting for contraceptive services.} \lrem{These moments were comparable to the ones raised by AGYW as those that make them anxious about approaching the pharmacists. These include, }\ladd{AGYW expressed that they have had negative experiences with pharmacists, such as }being scolded by the pharmacist for not knowing anything about contraceptives, being told that they are too young for sex, and being asked who their sexual partners are. \ladd{To avoid confrontation and uncomfortable interactions, AGYW rehearsed how to talk to the pharmacist so that they can avoid crying during counseling (AGYW-2), thought of ways to convince the pharmacist to maintain confidentiality (AGYW-2), and considered going to a pharmacist where the AGYW is not known to the pharmacist (AGYW-3).}

We highlight two topics that revealed disagreements among pharmacy staff. First, pharmacists expressed different sentiments with involving young men in contraceptive education. With respect to a screen on the app showing a female body (Figure \ref{fig:app} b), one pharmacy staff said, \textit{``If we put a picture of a man in the display here, we are trying to put this as a collective responsibility…So the moment you start omitting man here is the moment you walk this journey alone. But if you include the man here you will get a lot of support here and the impact will be great,”} indicating the benefits of educating a wider audience about contraception. However, some pharmacists said that the app should adhere to its original intent to focus on AGYWs' needs and preferences. 

Second, the young age of AGYW presented a challenge for some pharmacists, but not all, to provide contraceptive services. Pharmacists were concerned about being seen as a business that gave AGYW \textit{``a ticket to promiscuous-ity”} and being known as the \textit{``clinic that offers these services that makes our girls behave this way.”} However, not all pharmacists thought the client’s age swayed their decision to provide contraceptive services: \textit{``At the end of the day you end up giving what you will be giving. You end up exposing her to these questions, the methods. Whether she's 15 or 24, whether she's below 18… So is [age] really important?”}

Not all pharmacists shared the same beliefs, indicating the possibility of inconsistent practices. The variability of pharmacists’ attitudes seemed to contribute to AGYW’s anxiety and uncertainty about whether they will encounter a judgemental or friendly pharmacist. 

\subsubsection{Challenges in counseling}

In response to the 20 minute role play, observing pharmacy staff participants said that most AGYW were unwilling to stay for that long. \lrem{According to pharmacy staff, m}\ladd{M}any AGYW present themselves as \textit{``all-knowing,”} because they have been using a method for a while, know what they want from the pharmacist, and do not desire counseling \ladd{(Pharmacy-1)}. There are also AGYW who send others to request P-2 \ladd{(the brand name of an emergency contraceptive pill, Postinor-2)} on their behalf, resulting in challenges with properly counseling AGYW: \textit{``Some will even send boda boda \ladd{[motorcycle or bicycle taxi]} guys, they will send their younger siblings…the end users will not come, they will send someone else. So even to counsel them, it is not easy\lrem{.}”} \ladd{(Pharmacy-1).} Other times, while they are able to approach the pharmacists, \lrem{they}\ladd{AGYW} pretend to obtain contraceptives for someone else or hide their identities by wearing a hijab\ladd{, a head covering conventionally worn by Muslim women,} to cover their face \ladd{(AGYW-2)}. These deliberate choices that AGYW make for obtaining contraceptives present challenges for the pharmacy staff to counsel AGYW. \ladd{However, AGYW avoided interactions with pharmacists because they worried about the pharmacist’s demeanor towards AGYW, being seen by someone they know at the pharmacy, and the pharmacist not maintaining confidentiality. These concerns stemmed from AGYW’s prior negative interactions with pharmacists, such as pharmacists interrogating AGYW about why they were here, pharmacists saying loudly what the AGYW’s request is in a way that was audible to other clients, and pharmacists telling AGYW that they were too young for sex.}

\subsection{Differences between Older AGYW and Younger AGYW}

We explain differences between older AGYW and younger AGYW with respect to language, digital, and health literacy levels. We elaborate on differences in their perceptions of and interactions with the app. 
\subsubsection{Language literacy}

\ladd{While younger AGYW’s lower education levels were not unexpected, they influenced younger AGYW’s comfort with engaging in text-based content of the app.}\lrem{Younger AGYW’s education levels were a main reason for why they were less comfortable with reading compared to older AGYW, whose majority completed secondary school.} \ladd{Younger AGYW were overall less comfortable than older AGYW with respect to comprehending English. During workshop activities, younger AGYW used Kiswahili or a mixture of Kiswahili and English, while older AGYW were more comfortable conversing in English. Younger AGYW engaged with the app’s English content less often and instead relied on the Kiswahili content more. For both older and younger AGYW, DhoLuo was the least popular language option. We learned that DhoLuo is more commonly spoken by adults in Kisumu county, rather than the younger population.} Some younger AGYW preferred to only listen to audio and watch videos, saying that the audio helped them quickly understand the content, and noted their benefits for a non-literate individual. One participant said that the audio reminded her of listening to a teacher, which encouraged her to pay attention to the content more keenly than reading it (AGYW-2). A few younger AGYW used the audio feature without instruction or explanation from facilitators.

\subsubsection{Digital literacy}

Older AGYW used the tablets very comfortably, likely because of their prior experiences with smartphones. We did not notice any older AGYW requiring assistance with tablet or app usage. On the other hand, many younger AGYW were not familiar with touchscreen devices and were tapping parts of the screen that were irrelevant to the functionality of the app. As a result, facilitators assisted younger AGYW with app usage, such as with launching the app (i.e., tapping the app icon to open the app) and finding the volume adjustment buttons on the side of the tablet. This was expected given their lack of experience with and ownership of smartphones. One younger AGYW told us that while she found the audio feature helpful, her encounter with it was unintentional: \textit{``At first, I was just choosing randomly and I found it. Now I know how to find it”} (AGYW-2). This indicates that younger AGYW took an experimental approach to understanding how to use the app and tablet \ladd{as opposed to their interactions with the app and tablet being informed by prior experiences with technology}. 

\subsubsection{Health literacy}

Both older and younger AGYW expressed that the app debunked rumors and misinformation, indicating that both subpopulations were misinformed to some degree. However, older AGYW suggested that they were more familiar with the app’s information relative to younger AGYW. Older AGYW had prior experiences searching for contraceptive information and were mostly familiar with the contraceptive methods on the app. The app was helpful for confirming or augmenting prior knowledge with additional information, such as the administration of contraceptive methods. Some older AGYW were interested in \ladd{obtaining} in-depth information on specific contraceptive methods, such as injectables, as opposed to general contraceptive information that was on the app. Moreover, older AGYW expressed that the app provided them with the \textit{``reassurance [that] all will be well”} because it laid out all potential concerns \ladd{(AGYW-1) and the app’s content resonated with what the AGYW had experienced before (AGYW-3)}. For example, one participant said that the app’s information minimized their fear of side effects of injectables and gave them a better sense of the right method for them. Videos also contributed to feelings of reassurance \ladd{(AGYW-1)}. Provider-based videos provided additional information that boosted their confidence in decision-making and peer videos helped AGYW feel less isolated. 

On the other hand, younger AGYW expressed that much of the app’s information was novel to them. For example, one younger AGYW told us that they learned that only condoms could prevent HIV and STIs. Younger AGYW’s behaviors also suggested the novelty of the app’s information. They were writing notes on the app’s content, discussing among themselves about the content, and asking follow-up questions to the workshop facilitators about the side effects and the length of method effectiveness. Younger AGYW also discussed how their journeys in the pharmacy setting would be impacted by the app. They said that before the pharmacy visit, they would have limited knowledge on available methods, their use, benefits, and side effects. They would also be uncertain of how to engage with the pharmacist and express their needs, leading to high pressure. However, the app could allow AGYW to gain knowledge on contraception and feel more confident in expressing their needs to the pharmacist. These observations and responses imply that younger AGYW have less contraceptive knowledge to begin with as compared to older AGYW, but the app could allow them to gain new information and confidence in approaching the pharmacist.

\subsection{The Roles of the Pharmacist}
Pharmacists and AGYW agreed that pharmacists play a critical role in facilitating contraceptive education and decision-making. We elaborate on the different roles that the pharmacists might take on with the introduction of the app.

\subsubsection{Follow best counseling practices}

Pharmacists understood that AGYW avoided counseling due to fear of long questioning. One pharmacist said that \textit{``counseling and questioning are different,”} implying that counseling should not be an interrogation. Overall, pharmacists agreed on the importance of creating a safe space to allow AGYW to openly share their needs and concerns. Creating a safe space requires the pharmacist to place their personal beliefs aside: \textit{``whatever you are saying, don't judge them. Always let them have their way so that they can make the decision for their own. Don't decide for them”} \ladd{(Pharmacy-1)}. AGYW expressed that they want to be treated with respect and trust that pharmacists would maintain confidentiality. \ladd{They recommended providing pharmacy staff training so that they can better talk with AGYW (AGYW-3), suggesting that better counseling training could make AGYW feel more comfortable approaching the pharmacy setting.} Younger AGYW believed in the pharmacists’ professionalism to manage the app and were not concerned about the app being used in the pharmacy setting. 

\subsubsection{Accommodate different AGYW needs}

Older AGYW thought that counseling could occur at different moments of the pharmacy visit depending on the AGYW’s background. One older AGYW said that direct in-person counseling was more valuable than using the app for non-literate AGYW and another said that the app was more valuable for those who were more aware of contraception and had basic knowledge: \textit{``The ones who will use this app, they are aware. So for the ones who do not understand it need more consultation”} \ladd{(AGYW-1)}. However, younger AGYW did not question that they were a population of interest for the app. \ladd{While y}\lrem{Y}ounger AGYW \lrem{did }express\ladd{ed} that they might encounter issues with app usage, \lrem{but}\ladd{they} did not recognize this as a\lrem{n} \ladd{significant} issue because they thought that the pharmacist could \lrem{assist}\ladd{facilitate app use}. Pharmacists were willing to help AGYW, especially those who were not tech savvy, to use the app. Overall, these sentiments indicate AGYW’s perception that the pharmacist could support AGYW of varying language, digital, and health literacy levels. 

\subsubsection{Answer AGYW’s questions}

\ladd{In the workshops, younger AGYW asked follow-up questions about the side effects and the length of method effectiveness, suggesting that the app helped them structure questions.} Younger AGYW wanted the pharmacist to confirm the information they learned from the app and provide more explanations if needed\ladd{, suggesting that they trust the pharmacist’s expertise and guidance (AGYW-2)}. \lrem{Their behaviors suggested that they would need the pharmacist to answer the questions they structured after using the app.}Pharmacists also recognized that although the app provided answers to important questions about contraception, AGYW may not fully understand the concepts that were being addressed in the app, so \textit{``the service provider will be in a position to explain further…the terms [that AGYW do not understand]...`What are hormones and what do they do in our body, why will there be changes, and if I stop using them will there be other changes?'”} \ladd{(Pharmacy-1).} Pharmacists \lrem{were keen to \textit{``give [AGYW] room to ask questions,”}}\ladd{wanted to give room for AGYW to ask their questions at their own pace,} but recognized that they needed deep knowledge of contraception and have to learn to communicate with AGYW: \textit{``[pharmacists] need to be trained on the services and especially the methods, the side effects, the do’s and the don'ts\ladd{}\lrem{.}”} \ladd{(Pharmacy-1), echoing a similar sentiment that AGYW expressed on wanting the pharmacists to be trained on how to converse with AGYW (AGYW-3).} 

\subsubsection{Provide guidance on method choice}

The desire for the pharmacist to supportively \textit{``affirm your choice”} \ladd{(AGYW-1)} and provide guidance on appropriate methods was more strongly expressed by older AGYW than younger AGYW. Younger AGYW were more focused on thinking about how the pharmacist could support information provision as opposed to guidance on method choice. In discussing the benefits of making the app public, older AGYW wanted the pharmacist to facilitate the process of finalizing on a method choice, as summarized by this participant: \textit{``I feel like when you go to the hospital or pharmacy it's all about seeing the doctor…and tell him or her what's really happening or what I want. So I think the app should just be on the Play Store so that you can just have the information at hand. You can go and tell the doctor you want this and this\lrem{.}”} \ladd{(AGYW-1).} This AGYW wanted to spend more time discussing preferred method choices with the pharmacist than reviewing general contraceptive information with them. Older AGYW suggested that, if the app were available for use outside of the pharmacy setting, they would know what method they want by using the app beforehand and use the counseling time to request in-depth information about their preferred method or immediately request a transaction: \textit{``You can go for consultation but at least you know what you are going for. So maybe say you're going for depo, you need more explanation\lrem{.}”} \ladd{(AGYW-1).}  Older AGYW explained that they would be able to approach the pharmacist with less anxiety because the app would answer most of their questions prior to the pharmacy visit.

%% file: sections/06_discussion.tex
\section{Discussion}

Our main findings are as follows: 
\begin{enumerate}
\item Pharmacists’ interests, goals, and priorities can differ from those of AGYW, contributing to AGYW’s anxiety and avoidance of approaching the pharmacy setting for contraceptive services.
\item Older AGYW aged 18-24 years and younger AGYW aged 15-17 years present different language, digital, and health literacy levels, where younger AGYW were less educated and hence less comfortable with reading, less experienced with technology, and less informed about contraception and health overall. 
\item Pharmacists and AGYW validated the role of the pharmacist. Pharmacists will possibly take on multiple roles after the app is introduced in the pharmacy setting to accommodate different AGYW’s needs.
\end{enumerate}

Overall, the app received positive feedback from workshop participants. Videos, especially those featuring peers, were popular among AGYW, confirming the strengths of video-based education as established in prior work in HCI4D ~\cite{10.1145/2737856.2738023, 4937388, 10.1145/3544548.3581458, 10.1145/3491102.3501950}. Co-design allowed us to amplify AGYW’s experiences with contraception and pharmacies and incorporate their voices into iterations of the app ~\cite{10.1145/3544548.3581458, 10.1145/3613904.3642532, 10.1145/3283458.3283476}.

In this section, we focus on three components: 1) potential role of intermediation in the pharmacy setting; 2) potential differential impact of LMIC-based interventions on subgroups; and 3) challenges and opportunities for technological interventions in LMIC healthcare settings.

\subsection{Intermediation in the Pharmacy Setting}
We first clarify why intermediation is an appropriate lens of analysis. The physical and human infrastructure of the pharmacy setting enables an environment where technology can be supported ~\cite{10.1145/1753326.1753718, 1626204}. The pharmacy setting affords a private room for AGYW to learn about contraception, provides a tablet that eliminates requirements of smartphone ownership, and has a pharmacist who can facilitate contraceptive education and decision-making. Similar to other ICTD/HCI4D interventions ~\cite{4937388, 10.1145/2737856.2738023}, a central motivation of this intervention is to provide complementary expertise to that of the pharmacist. In some ways, the app may have more information than what the pharmacist can provide to AGYW, given known limitations in counseling time and pharmacy staff knowledge ~\cite{gonsalves2020pharmacists, gonsalves2019regulating}. Potential intermediation in the pharmacy setting resembles those observed in LMICs where the intermediary takes on tasks in addition to their primary task ~\cite{10.1145/3613904.3642099, 10.1145/3449118, 10.1145/2909609.2909655}. With the introduction of the app, pharmacists are expected to not only help AGYW with achieving the primary goal of their pharmacy visit – to obtain contraceptive services – but also app usage. We anticipated AGYW's lack of interest in the app since it is not their main purpose for visiting the pharmacy and addressed this concern by adding interactive multimedia and creating an appealing user interface.

Next, we consider how intermediated interactions can contribute towards a successful intervention in the pharmacy setting, adding to literature that recognizes the importance of human relationships for a technological endeavor ~\cite{10.1145/2369220.2369258, 1626204, 10.1145/3613904.3642099}. 

\subsubsection{The differential role of the pharmacist}
One of our biggest learnings was the differences between older AGYW and younger AGYW. It was not feasible for us to create two apps and we wanted to make a single, robust app that fits both subpopulations. Because of the limitations of a single app, the human agent - the pharmacist - is critical for providing the flexibility to accommodate a broad range of AGYW’s needs and behaviors ~\cite{10.1145/2369220.2369258, 10.1145/3173574.3174213}. For older AGYW who had basic contraceptive knowledge, the primary benefit of the app was the reassurance it provided in their decision-making process. The app was not strictly viewed as an information provision tool, but also as a decision aid tool for evaluating contraceptive options. \ladd{Some older AGYW were more interested in obtaining in-depth information on specific contraceptive methods of interest, suggesting that they already had an idea of which contraceptive method was right for them and that they had topics in mind that they were wanting to learn about. The app provides more general information than what these older AGYW want. As such, the pharmacist will likely be relied upon to discuss topics of interest.} \textbf{For older AGYW, the pharmacist could be a ``last-mile connector” ~\cite{10.1145/2369220.2369258} who provides guidance on making informed choices or facilitates the last step in obtaining a contraceptive method (RQ1)}. 
 
On the other hand, for younger AGYW who were less technically experienced and \ladd{less} informed about contraception, pharmacists may need to play a more active role throughout the pharmacy visit. To support successful technology use, pharmacists may need to practice ``proximate enabling” ~\cite{10.1145/1753326.1753718} by helping AGYW with using the app. For younger AGYW who had nascent contraceptive knowledge, the app helped build foundational knowledge that is necessary for reaching the point of making informed choices. While the app exposed AGYW to standard information, it did not answer all of their questions, as \lrem{became}\ladd{was} evident \ladd{in the workshops} when younger AGYW asked questions \ladd{about side effects and length of method effectiveness}\lrem{ in the workshops} after using the app. \ladd{Pharmacy staff also expected that AGYW might ask questions about difficult health concepts, such as hormones, that the app does not explain.} To explain \lrem{difficult health concepts, such as hormones,}\ladd{health information} in understandable language, pharmacists may need to practice ``proximate translation,” which is the simplification of information ~\cite{10.1145/1753326.1753718}. \textbf{The pharmacist plays a critical role in bridging younger AGYW’s knowledge gap to build a foundation for making informed choices (RQ1)}. 

Pharmacists’ ability to execute various roles, however, could depend on their domain knowledge. Pharmacists may find that communicating with younger AGYW about contraceptive information is easier than providing nuanced guidance on method choice for older AGYW. However, the ease of communication with younger AGYW could depend on pharmacists’ comprehensive knowledge of contraception and even sexual and reproductive health holistically, as younger AGYW are likely to raise broader health questions. In this sense, pharmacists might struggle less in communicating with older AGYW who want a quick conversation to have their method choice validated by the pharmacist. However, older AGYW may want more method-specific information that was not addressed with the app, which would require pharmacists to have in-depth knowledge of particular contraceptive methods.

The following sections build on the strengths of intermediation, as described above, to discuss important considerations for technological interventions in intermediated scenarios in LMICs.
 
\subsection{Extending Impact to a Hard-to-Reach Population}

Especially in LMICs where technology use is growing, education levels vary, and access to resources are limited, it is important to consider who is being reached with interventions but also those who fall just outside their reach. We draw on a conversation in ICTD to illustrate how intervention decisions can include or exclude certain populations. Despite the strong presence of feature phone owners in LMICs ~\cite{Silver_Johnson_2018, Kenya_2023, Tanzania_2015, Uganda_2022}, smartphone users became a primary interest for ICTD projects as they became technologically eas\lrem{ier}\ladd{y}-to-reach due to growth in smartphone penetration in LMICs ~\cite{Delaporte_Bahia_2022} and the perception that smartphones are innovative, convenient, and desirable. However, some ICTD researchers advocate for the continued support of feature phone users ~\cite{10.1145/3613904.3642099, 10.1145/3555648, doi:10.1177/2050157918776684}. 

We extend this reflection to our work to understand how intervention decisions can differently impact older AGYW who are technologically eas\lrem{ier}\ladd{y}-to-reach and younger AGYW who are hard\lrem{er}-to-reach. Older AGYW’s access to and experiences with technology in addition to their contraceptive knowledge likely contributed to them thinking that the app would be irrelevant to AGYW who were non-literate or lacked basic contraceptive knowledge. They also requested more app features and wanted the app publicly available so that smartphone users, like themselves, can use it before \lrem{their }visiting the pharmacy setting. While changes made according to these suggestions could encourage wider use of the app by eas\lrem{ier}\ladd{y}-to-reach\ladd{, older} AGYW who can access and use technology, \ladd{as explained earlier, the app needs to be studied and refined before it is ready to be made widely available. AGYW who have access to smartphones might share their devices with family members; an app installed on a shared device risks exposing AGYW, which could harm their safety. Moreover,} \textbf{over focusing on a technologically eas\lrem{ier}\ladd{y}-to-reach population could hinder efforts to reach a hard\lrem{er}-to-reach population with less resources (RQ2)}. \ladd{If the app were to be designed exclusively for older AGYW, younger AGYW, who are arguably the subpopulation who has a relatively larger knowledge gap, may not have other opportunities to learn about contraception. As such, we} \lrem{We }prioritize building a single, robust app that \lrem{can }extend\ladd{s} impact to AGYW who are hard\lrem{er}-to-reach because of their low levels of education, technology experience, and maturity around sexual and reproductive health. Not only does the pharmacy setting provide an environment where technology is accessible, but the intermediation of the pharmacist could extend the benefits of the app to a wide range of users ~\cite{10.1145/1753326.1753718}. For hard\lrem{er}-to-reach AGYW, the benefits of technology may only be realized with intermediation in the pharmacy setting ~\cite{10.1145/2369220.2369258}. 


\subsection{Implications for Complex Healthcare Settings in LMICs}
There have been numerous efforts in HCI to understand how technologies fit into complicated and sensitive domains, including within health ~\cite{10.1145/3613904.3642245, 10.1145/3637323, 10.1145/3411764.3445410, 10.1145/3313831.3376465}. We contribute to these efforts by reflecting on how \lrem{our}\ladd{the Mara Divas} app could integrate into a social system with human relations, beliefs, and practices and a medical system with a standard for understanding contraception and improving health outcomes. In doing so, we discuss implications for technological interventions in LMIC-based healthcare settings with hopes to help HCI researchers in the health domain \lrem{identify affordances that can make }\ladd{clarify how} technology \ladd{can be} beneficial and impactful. \ladd{These implications and considerations are based on our understanding of the social and cultural contexts of this work and experiences with designing, developing, and getting feedback on an initial app prototype.}

Healthcare providers in LMICs might lack adequate training or motivation ~\cite{scott2016non, bitton2017primary}, possibly because healthcare settings are overburdened. Our intervention considered this concern by creating an app that in itself does not depend on pharmacy staff’s knowledge but rather aims to augment it. However, the app could complicate information penetration. In one way, the app could promote views that are difficult for the pharmacist to endorse. In another way, the app could inform AGYW about difficult topics that the pharmacist could not represent. \ladd{For example, in the pharmacy staff workshop role play, the pharmacist asked the AGYW if religion allowed her to use contraception, implying the influence of religion on a pharmacist’s stance on contraceptive service provision. It is plausible that a pharmacist might not encourage AGYW to use contraception. The app, on the other hand, provides evidence-based information that could help AGYW in meeting their goals to prevent pregnancy. However, the potential tension between pharmacists’ views and the app’s information risks exposing AGYW to contradicting views.} \lrem{To minimize the risk of exposing AGYW to contradicting views, the messaging that is promoted in the intervention and that represented by providers should be aligned. }\textbf{Standardization of health messaging is crucial because messaging could promote accurate and consistent information dissemination and penetration and impact behavior change and health outcomes ~\cite{agha2010intentions, mwaikambo2011works} (RQ2)}. However, achieving standardization in LMICs is challenging because of issues with inadequate funding, overburdened healthcare workers, and healthcare worker complacency ~\cite{adovor2021medical}. In commercial settings, such as pharmacies, profit-maximizing motives ~\cite{wulandari2021prevalence, miller2016performance} and the lack of monitoring and enforcement of rules ~\cite{wulandari2021prevalence} could pose additional barriers for ensuring that service providers adhere to standardized messaging. Moreover, the controversy surrounding sensitive topics including contraception complicates the process of standardization and acceptance of a chosen standard.

Client-provider relations can impact clients’ willingness to seek healthcare services and reception towards technological applications in healthcare ~\cite{10.1145/3613904.3642245}. Our population of interest felt discouraged from approaching pharmacies because of misaligned incentives, goals, and priorities with the pharmacist. \ladd{AGYW were anxious about interactions with the pharmacists from before their visit, causing them to rehearse the interactions, wear cover-ups to hide their identities, or avoid the pharmacy setting entirely.} \lrem{It is possible that s}\ladd{S}trained AGYW-pharmacist relationships could \ladd{consequently} dissuade AGYW from approaching the pharmacies to use the app. This \lrem{case}\ladd{concern} demonstrates the limitations of technology alone to make an impact. \textbf{While relationship-building was not intended to be a focus of our technological intervention, it may be necessitated as part of broader motivation to improve AGYW-centered contraceptive resources and contribute to improved health outcomes (RQ2)}. To realize the potential significance of intermediated interactions in the pharmacy setting, pharmacists will need to improve their relationships with AGYW, a critical foundation of intermediated interactions ~\cite{10.1145/2369220.2369258}, by building their competencies and attitudes to promote an AGYW-friendly environment ~\cite{chandra2014contraception}. 

The physical infrastructure of commercial healthcare settings, which tend to lack waiting areas and private counseling areas ~\cite{de2018adolescent}, raises unique considerations for interventions in these settings. \ladd{In deciding to place the app in the pharmacy setting, we also considered the privacy risks of using the app at the pharmacy counter where it is highly visible and open. As such, we}\lrem{ We} intend for the \ladd{Mara Divas} app to be used only in pharmacies with private counseling rooms to provide a safe space to learn and discuss a sensitive topic. The provision of a private room was also a motivation for targeting tablets for our app. \lrem{Requiring AGYW to use the app at the pharmacy counter where it could be visible to people would impose a higher privacy risk for AGYW than using the app in a private back room.}\textbf{The physical infrastructure of healthcare settings in LMICs, which is often restricted and overcrowded, is an important aspect of intervention design that, if not considered carefully, could harbor risks (RQ2)}.

\subsection{Limitations and Future Work}

We acknowledge that sampling in the workshops could bias the design of the app, which may not benefit all AGYW. We plan to address this by further exploring more AGYWs’ experiences with the app and in the pharmacy setting in the feasibility study, which took place between July - September 2024, where approximately 100 AGYW engaged with the app and provided feedback \ladd{and pharmacy staff provided their perspectives on the app intervention}. The goal of the feasibility study was to evaluate the feasibility, acceptability, and appropriateness of the app in the pharmacy setting from the perspectives of AGYW and pharmacy staff at study sites. Findings from the feasibility study will be presented in a future paper.

This study was carried out in a single county in Kenya and its findings are not generalizable to the whole of Kenya. However, we believe that the findings from this study and the feasibility study will be useful for informing the next phase of research, which is to refine the app with consideration for uses in rural areas and other regions of Kenya. Our team plans a hybrid effectiveness implementation trial to concurrently examine both clinical effectiveness and implementation outcomes in the intermediated pharmacy setting. This research will include ongoing refinement of the app and strategies for integrating it into the pharmacy setting, as well as a cluster-randomized trial to examine clinical outcomes. 
 
\ladd{At this stage of the research process, we think it is appropriate to optimize situating the app into the pharmacy setting where AGYW already seek contraceptive services. After the value and utility of the app in the pharmacy setting is studied and validated, there may be opportunities to extend the benefits of this app beyond the pharmacy setting. This process, though, will entail its own strategy and design changes to the app to ensure the app’s success as part of a wider contraception awareness agenda. Peer mobilizers, who were leveraged to recruit AGYW participants for the co-design workshops, could play an important role in helping peers prepare for pharmacy visits. For example, peer mobilizers with smartphones could assist AGYW without smartphone access in using the app before visiting the pharmacy setting. This process would entail training peer mobilizers to facilitate app usage, including smartphone usage. Future work could also involve the expansion of the population of interest for male peers by presenting male contraceptive methods in the app and expanding app usage to young men. Enabling wider availability of the app so that it can be used outside of the pharmacy setting could provide opportunities for AGYW and their male peers to use the app together and collaboratively learn about contraception.}

%% file: sections/07_conclusion.tex
\section{Conclusion}

AGYW in Kenya and more broadly in sub-Saharan Africa lack contraceptive decision-support resources that are tailored to this population. To contribute to efforts to empower an underserved population and improve health outcomes, we developed a person-centered, tablet-based app \ladd{- the Mara Divas app -} that provides contraceptive education and decision-support for AGYW in the pharmacy setting in Kenya. Through co-design workshops, we gathered app feedback and explored how our intervention could integrate into the pharmacy setting. Our findings suggest that our intervention will have a differential impact on AGYW depending on their age, education, technology experience, and contraceptive knowledge. We highlight the significant role that intermediation could play in the context of the app, with the pharmacist taking on various roles to accommodate different AGYWs’ needs.

More broadly, this work presents important considerations to make for technological interventions in intermediated healthcare scenarios in LMICs. We reflect on who is being reached and who is unable to be reached with these interventions and advocate for extending impact to a hard\lrem{er}-to-reach population that is characterized by their low levels of education, health literacy, and access to and familiarity with technology. We also recognize that effectively and meaningfully designing technology within healthcare settings in LMICs warrants a reflection on the complex system it is being designed for, including the medical system, human relations, and physical infrastructure. We hope that our work contributes towards a holistic understanding of the challenges and opportunities in LMIC-based interventions.

%% file: sections/08_appendix.tex
\section{Pre-Workshop Questionnaire} \label{question}
\label{appendix}

\subsection{AGYW}
\begin{enumerate}
    \item Demographic information
    \begin{enumerate}
        \item What is your date of birth?
        \item Age in years for AGYW
        \item What is your religion?
        \begin{enumerate}
            \item Christian
            \item Muslim
            \item African traditional
            \item Other:
        \end{enumerate}
        \item What is your highest level of education?
        \begin{enumerate}
            \item No schooling
            \item Primary school, not complete
            \item Primary school, complete
            \item Secondary school, not complete
            \item Secondary school, complete
            \item Post-secondary school
        \end{enumerate}
        \item Are you still attending school?
        \begin{enumerate}
            \item Yes
            \item No
        \end{enumerate}
            \item In the past 12 months, did you earn your own income?
        \begin{enumerate}
            \item Yes
            \item No
        \end{enumerate}
        \item Who do you live with? (Select all that apply)
        \begin{enumerate}
            \item No one (lives alone)
            \item Parent(s)/guardian
            \item Other family members (sibling(s), extended family)
            \item Boyfriend/partner/husband
            \item Friend
            \item Other (specify)
        \end{enumerate}
        \item Who provides financial support for you? (Select all that apply)
        \begin{enumerate}
            \item Self
            \item Parent(s)/guardian
            \item Other family members (sibling(s), extended family)
            \item Boyfriend/partner/husband
            \item Friend
            \item Other (specify)
        \end{enumerate}
    \end{enumerate}

    \item Mobile phone and app use
    \begin{enumerate}
        \item Do you have access to a mobile phone?
            \begin{enumerate}
                \item Yes
                \item No
            \end{enumerate}            
        \item Do you own the phone or is it shared?
            \begin{enumerate}
                \item Own
                \item Shared
            \end{enumerate}  
        \item With whom do you share the phone?
        \begin{enumerate}
            \item Parent
            \item Boyfriend/partner
            \item Other family member
            \item Other
        \end{enumerate}
        \item What kind of phone is it?
            \begin{enumerate}
                \item Smartphone
                \item Not a smartphone
            \end{enumerate}  
        \item How often do you use mobile devices (like a phone or tablet) in your daily life?
        \begin{enumerate}
            \item Multiple times a day
            \item At least once a day
            \item At least once a week
            \item Rarely
        \end{enumerate}
        \item Do you have your own mPesa line?
            \begin{enumerate}
                \item Yes
                \item No
            \end{enumerate}  
        \item Do you have a social media account or accounts?
            \begin{enumerate}
                \item Yes
                \item No
            \end{enumerate}  
        \item What mobile applications have you used? (if none go the next section)
        \begin{enumerate}
            \item Facebook
            \item Instagram
            \item WhatsApp
            \item TikTok
            \item mPesa
            \item Other (please specify)
            \item None - has never used a mobile app
        \end{enumerate}
        \item What is your level of comfort using mobile apps?
        \begin{enumerate}
            \item Extremely comfortable
            \item Comfortable
            \item Neutral
            \item Not very comfortable
            \item Not at all comfortable
        \end{enumerate}
    \end{enumerate}

    \item Relationships and sex
    \begin{enumerate}
        \item Are you currently in a relationship with a partner or partners?
            \begin{enumerate}
                \item Yes
                \item No
                \item No answer
            \end{enumerate}  
        \item How long have you been in your current relationship? (Months)
        \begin{enumerate}
            \item Less than 3 months
            \item 3-6 months
            \item 7-12 months
            \item 13-24 months
            \item More than 24 months
        \end{enumerate}
        \item Do you know your partner's age?
            \begin{enumerate}
                \item Yes
                \item No
                \item I choose not to answer
            \end{enumerate}  
        \item How old is your partner?
        \item Are you currently living with that boyfriend/partner?
            \begin{enumerate}
                \item Yes
                \item No
                \item No answer
            \end{enumerate}  
        \item Does your boyfriend/partner provide you with any financial support or things you need?
            \begin{enumerate}
                \item Yes
                \item No
                \item No answer
            \end{enumerate}  
        \item What is your current marital status?
        \begin{enumerate}
            \item Never married
            \item Married
            \item Separated
            \item Divorced
            \item Widowed
        \end{enumerate}
    \end{enumerate}

    \item Reproductive history, contraceptive experience/desire
    \begin{enumerate}
        \item How many times in your life have you been pregnant? (number) (If answer is non-zero, ask questions next three questions. Otherwise, skip to fourth question.)
        \item How many times have you given birth? (number)
        \item Have you ever done anything to end a pregnancy (induced abortion)?
            \begin{enumerate}
                \item Yes
                \item No
                \item I choose not to answer
            \end{enumerate}  
        \item How many times have you attempted an abortion?
        \item Have you ever taken a home pregnancy test?
            \begin{enumerate}
                \item Yes
                \item No
                \item No answer
            \end{enumerate}  
        \item Have you or a sexual partner ever used any of the following other methods to prevent pregnancy? (Select all that apply)
        \begin{enumerate}
            \item Withdrawal (pulling out)
            \item Natural family planning (rhythm or calendar method)
            \item Birth control pill (the Pill)
            \item "E" pill ("EC" or emergency contraception)
            \item Shots or injections (Depo)
            \item IUCD (Intrauterine Device, coil, IUD)
            \item Implant (Jadelle, Implanon, Nexplanon)
            \item Tubal ligation, vasectomy or sterilization (getting tubes tied)
            \item Diaphragm
            \item Condoms
            \item Other - please specify:
            \item I have never had vaginal sex
            \item I have never used any
        \end{enumerate}
        \item In the last 3 months have you or a sexual partner used any of the following other methods to prevent pregnancy? Please include methods you were taking or using, like the Pill or the IUD, even if you did not have sex.
        \begin{enumerate}
            \item Withdrawal (pulling out)
            \item Natural family planning (rhythm or calendar method)
            \item Birth control pill (the Pill)
            \item "E" pill ("EC" or emergency contraception)
            \item Shots or injections (Depo)
            \item IUCD (Intrauterine Device, coil, IUD)
            \item Implant (Jadelle, Implanon, Nexplanon)
            \item Tubal ligation, vasectomy or sterilization (getting tubes tied)
            \item Diaphragm
            \item Condoms
            \item Other - please specify:
            \item I have never had vaginal sex
            \item I have never used any
        \end{enumerate}  
        \item Have you continued using (method(s) above) until now?
            \begin{enumerate}
                \item Yes
                \item No
                \item No answer
            \end{enumerate}  
        \item How confident are you that the method(s) you are using or last used is right for you?
        \begin{enumerate}
            \item Not at all confident
            \item Slightly confident
            \item Somewhat confident
            \item Quite confident
            \item Extremely confident
        \end{enumerate}
        \item What method(s) of family planning did you purchase at a pharmacy in the last 3 months? (select all that apply)
        \begin{enumerate}
            \item Birth control pill (the Pill)
            \item "E" pill ("EC", P2, emergency contraception)
            \item Shots or injections (Depo)
            \item IUCD (Intrauterine Device, coil, IUD)
            \item Implant (Jadelle, Implanon, Nexplanon)
            \item Diaphragm
            \item Condoms
            \item Other - please specify:
        \end{enumerate}
        \item Why did you choose the pharmacy as your source of family planning? (select all that apply)
        \begin{enumerate}
            \item Convenience
            \item Hours
            \item Privacy
            \item Cost
            \item Trust in provider
            \item Prior negative experience in a clinic
            \item Other (specify)
        \end{enumerate}
        \item In the last 3 months, have you had sex one or more times without any method to prevent pregnancy?
            \begin{enumerate}
                \item Yes
                \item No
                \item No answer
            \end{enumerate}  
        \item If you could use any birth control method you wanted, what method(s) would you use? (select all that apply)
            \begin{enumerate}
                \item I'm using the method I want
                \item Withdrawal (pulling out)
                \item Natural family planning (rhythm or calendar method)
                \item Birth control pill (the Pill)
                \item "E" pill ("EC" or emergency contraception)
                \item Shots or injections (Depo)
                \item IUCD (Intrauterine Device, coil, IUD)
                \item Implant (Jadelle, Implanon, Nexplanon)
                \item Tubal ligation, vasectomy or sterilization (getting tubes tied)
                \item Diaphragm
                \item Condoms
                \item Other - please specify:
                \item I don't want to use any method
                \item I do not know
            \end{enumerate} 
        \item Why are you not currently using the birth control method that you selected in the previous question? (select all that apply)
        \begin{enumerate}
            \item I can't afford it
            \item I don't know where I can get it
            \item It's too difficult to get it
            \item A health provider advised against it
            \item I don't like the side effects
            \item I'm not currently sexually active
            \item I'm trying to get pregnant
            \item My parents don't know I'm sexually active
            \item My partner doesn't want me to use it
            \item Other - please specify:
        \end{enumerate}
    \end{enumerate}
    
\end{enumerate}

\subsection{Pharmacy Staff}
\begin{enumerate}
    \item What is your age (years)?
    \item Gender of the pharmacy staff
        \begin{enumerate}
            \item Male
            \item {Female}
        \end{enumerate}
    \item What is your highest education level?
        \begin{enumerate}
            \item Primary incomplete
            \item Primary complete
            \item Secondary complete
            \item Technical college/trade school
            \item Diploma
            \item Bachelors/degree
            \item Doctorate
            \item Other (specify)
        \end{enumerate}
    \item What is your professional title?
        \begin{enumerate}
            \item Pharmacy technician
            \item Pharmacist
            \item Pharmacy owner
            \item Pharmacy assistant
            \item Other (specify)
        \end{enumerate}
    \item How long have you been working in the pharmacy setting?
        \begin{enumerate}
            \item Less than 1 year
            \item 1-2 years
            \item 2-4 years
            \item 5 or more years
        \end{enumerate}
    \item How confident do you feel advising AGYW on family planning options?
        \begin{enumerate}
            \item Not at all confident
            \item Slightly confident
            \item Somewhat confident
            \item Quite confident
            \item Extremely confident
        \end{enumerate}
    \item How confident do you feel advising AGYW on how to use family planning methods?
        \begin{enumerate}
            \item Not at all confident
            \item Slightly confident
            \item Somewhat confident
            \item Quite confident
            \item Extremely confident
        \end{enumerate}
    \item How confident do you feel advising AGYW on family planning side effects?
        \begin{enumerate}
            \item Not at all confident
            \item Slightly confident
            \item Somewhat confident
            \item Quite confident
            \item Extremely confident
        \end{enumerate}
\end{enumerate}

\section{Enlarged Images from Figure 1 and Figure 3} \label{big_images}
\begin{figure*}[hbt!]
\centering
\includegraphics[width=0.35\linewidth]{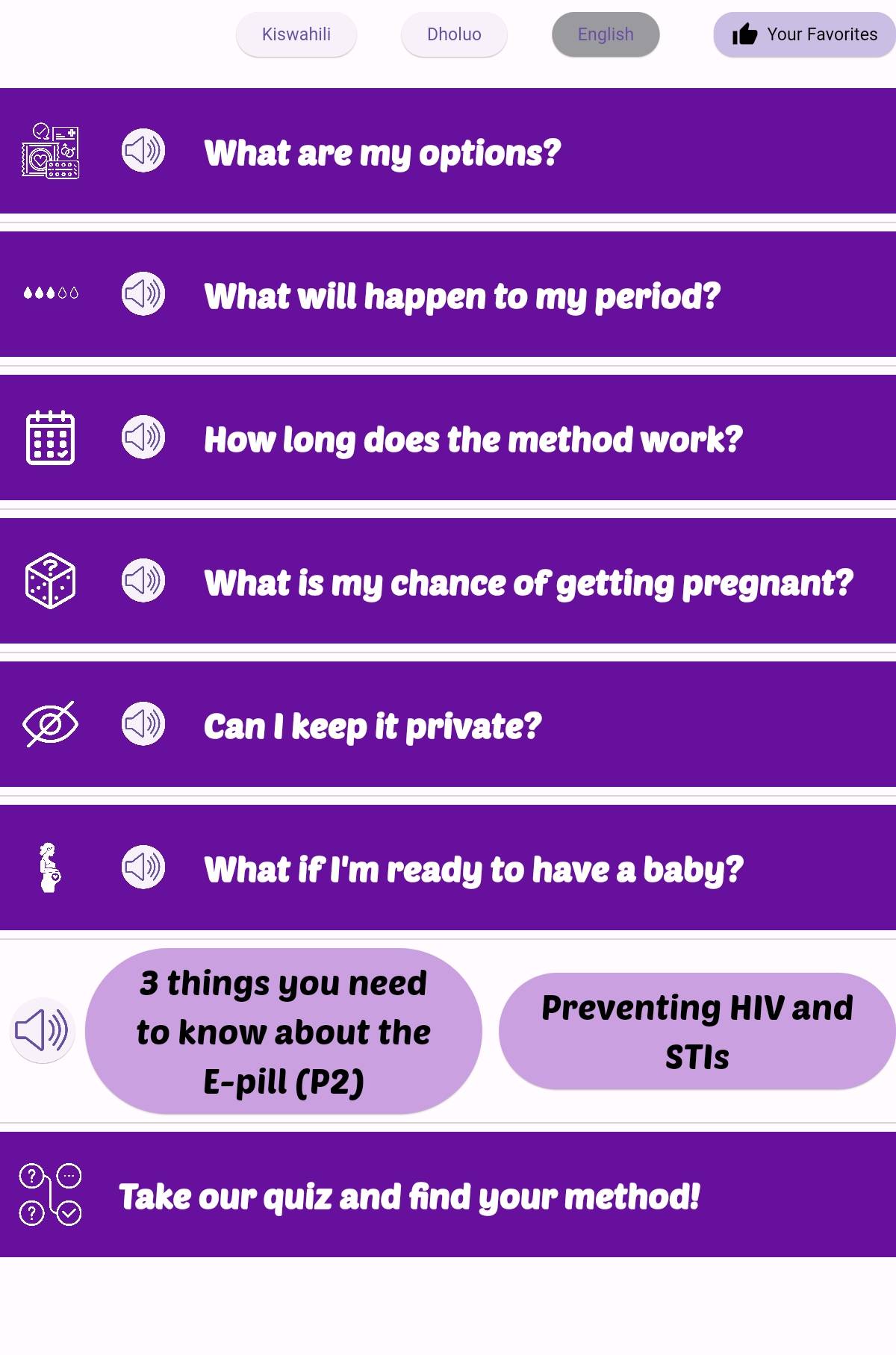}
\caption{Main screen with questions and topics that are of most interest and importance to AGYW.}
\Description{Main screen with nine questions and topics that are of most interest and importance to AGYW. Top of the screen has three language toggle buttons labeled ``Kiswahili,” ``Dholuo,” and ``English” where ``English” is grayed out to indicate its selection. These buttons are presented in all of the screens explained after. Top of the screen also includes a ``Your Favorites” button with a thumbs-up sign preceding the text, which is also in screens (b) and (e). The screen is filled with buttons, labeled in order from top to button, ``What are my options?” ``What will happen to my period?” ``How long does the method work?” `1`What is my chance of getting pregnant?” ``Can I keep it private?” ``What if I’m ready to have a baby?” and ``Take our quiz and find your method!” Each button is accompanied by a representative image of that topic and an audio button. Two buttons labeled ``3 things you need to know about the E-pill (P2)” and ``Preventing HIV and STIs” is also exhibited.}
\end{figure*}

\begin{figure*}[hbt!]
\centering
\includegraphics[width=0.35\linewidth]{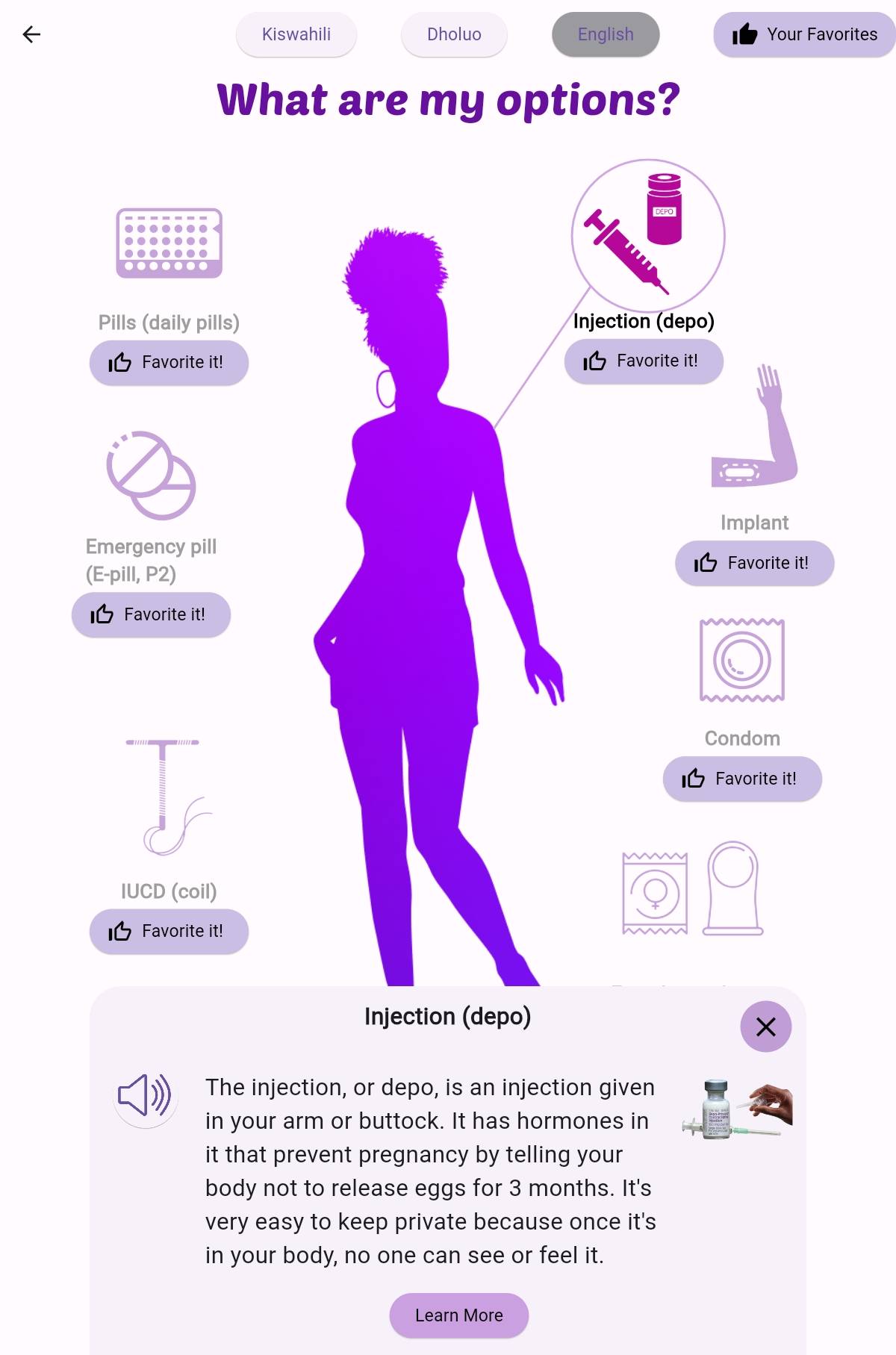}
\caption{Short explanations of each contraceptive method and where it is administered on a female body.}
\Description{Top of the page is titled ``What are my options?” A female body in the center of the screen with contraceptive method icons around it. On the left side of the body are icons and accompanying labels of pills (daily pills), emergency pill (e-pill, P2), IUCD (coil), and on the right side are injection (depo), implant, condom, and female condom. Below each label is a button ``Favorite it!” with a thumbs-up sign preceding the text. The injection (depo) is selected, indicated by a different color. A pop-up at the bottom of the screen shows a description of the injection (depo) with an audio button and a real image of the injection.}
\end{figure*}

\begin{figure*}[hbt!]
\centering
\includegraphics[width=0.35\linewidth]{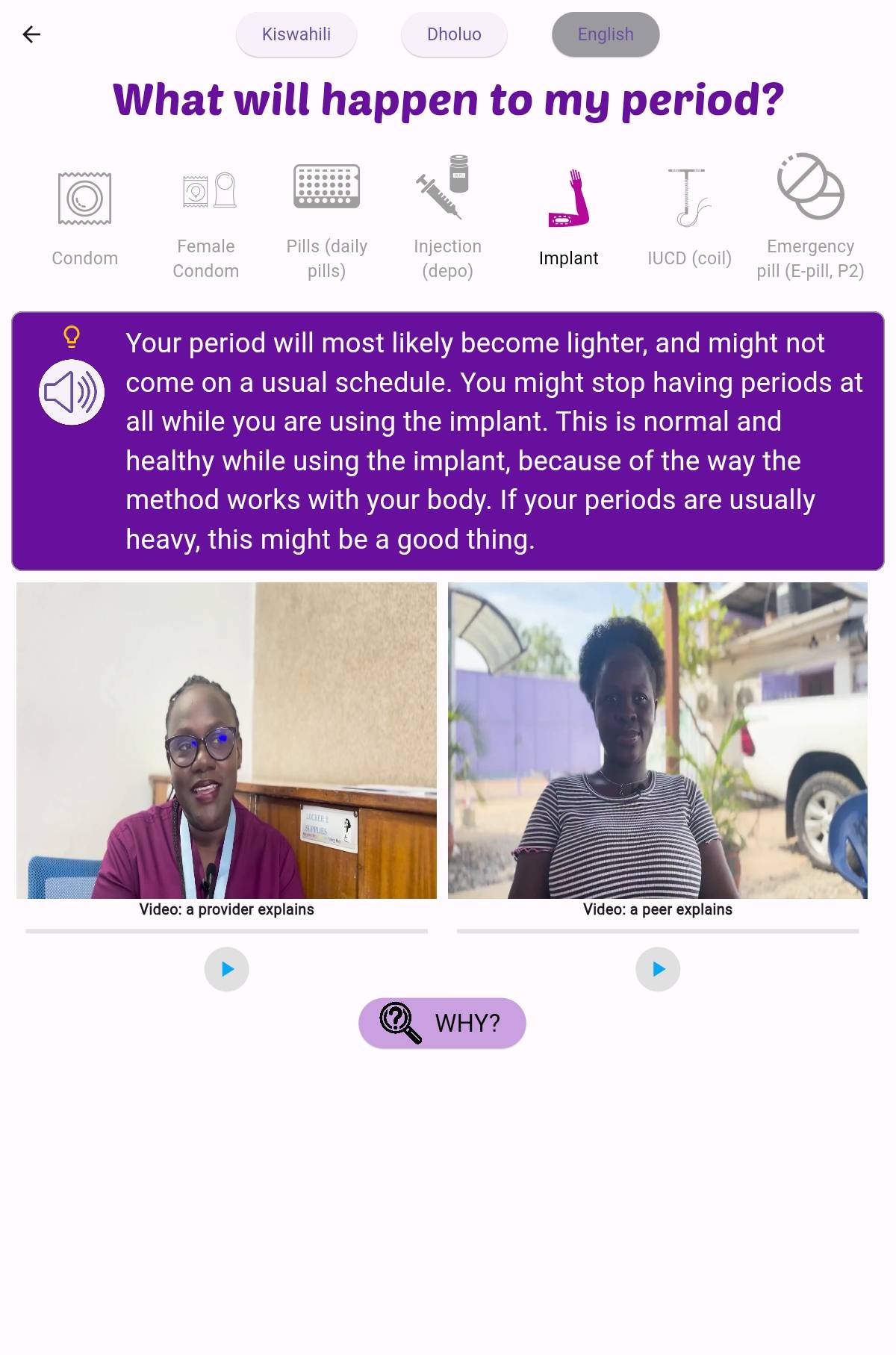}
\caption{Screen with text, audio, and videos (provider and peer).}
\Description{Top of the page is titled ``What will happen to my period?” Below the title are seven contraceptive methods with icons and labels, as described in screen (b). Implant is selected, indicated by a different color. Middle of the screen displays a text description of the method and an audio button. Below this is two videos - one of a provider, one of an AGYW peer - placed side by side. Below the videos is a button labeled ``WHY?” with a magnifying glass icon preceding the text.}
\end{figure*}

\begin{figure*}[hbt!]
\centering
\includegraphics[width=0.35\linewidth]{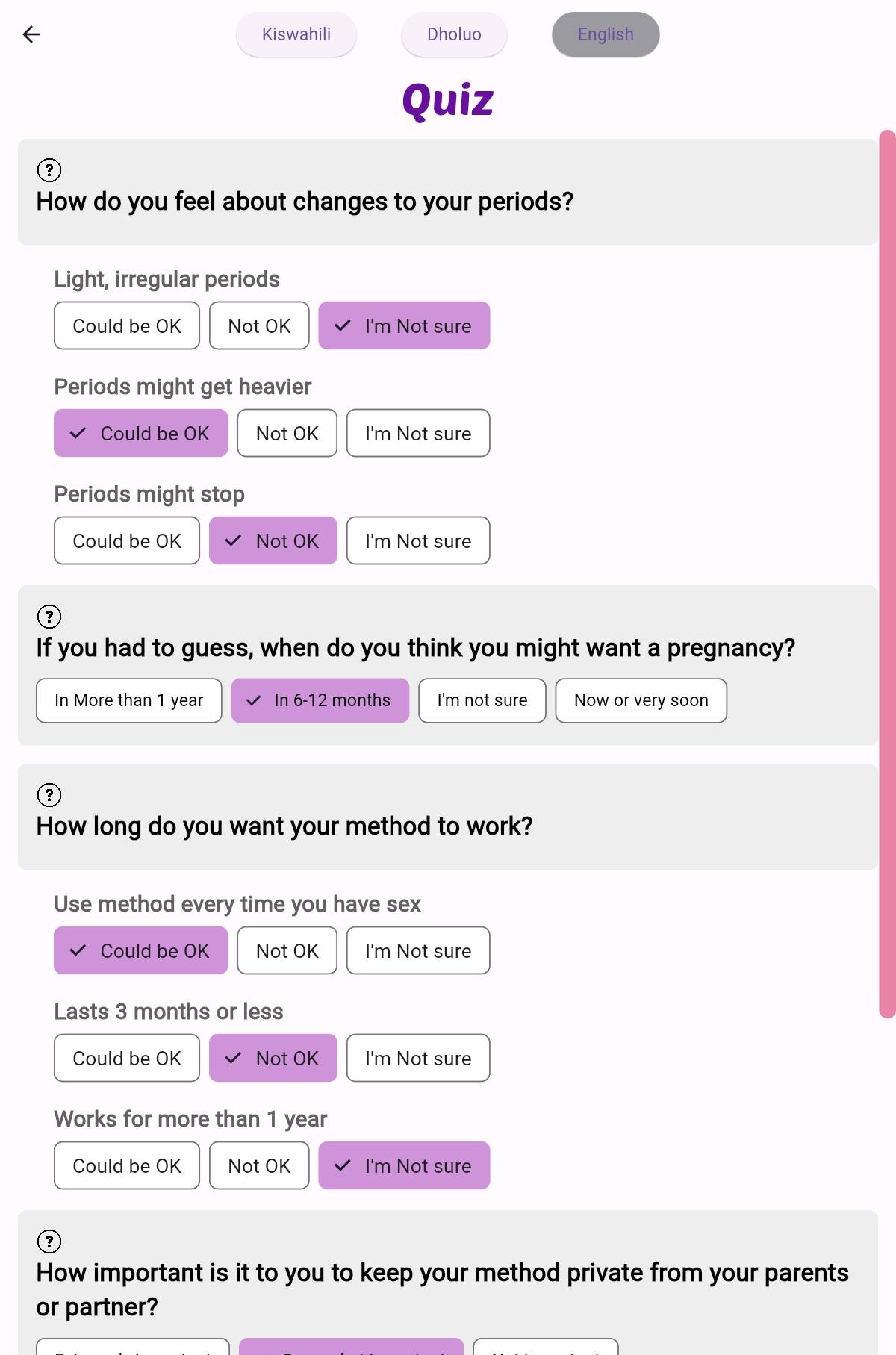}
\caption{Survey where users can indicate preferences for method characteristics. }
\Description{Top of the page is titled ``Quiz.” Multiple choice quiz/survey questions displayed are ``How do you feel about changes to your periods?” ``If you had to guess, when do you think you might want a pregnancy?” ``How long do you want your method to work?” and ``How important is it to you to keep your method private from your parents or partner?” Selections for each question are indicated by a different color. A scroll bar indicates continuation of the quiz.}
\end{figure*}

\begin{figure*}[hbt!]
\centering
\includegraphics[width=0.35\linewidth]{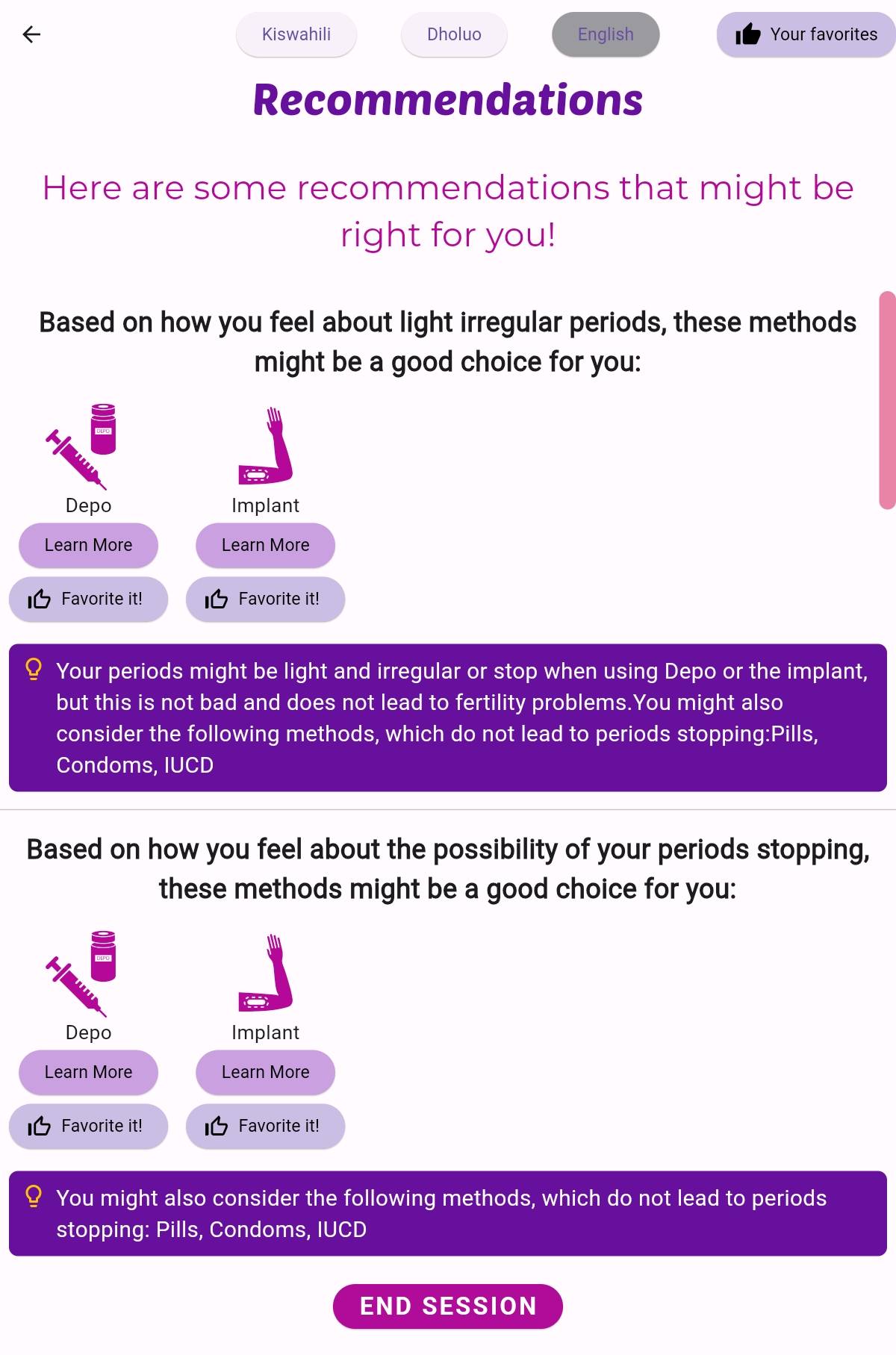}
\caption{Recommendations for contraceptive methods based on survey results.}
\Description{Top of the page is titled ``Recommendations” and below it, ``Here are some recommendations that might be right for you!” Recommendations displayed are titled ``Based on how you feel about light, irregular periods, these methods might be a good choice for you:” and ``Based on how you feel about the possibility of your periods stopping, these methods might be a good choice for you:” with icons and labels of the depo and implant. Below each method label is a button labeled ``Learn More” and ``Favorite it!” with a thumbs-up sign preceding the text. Below the buttons are short relevant explanations for other recommendations. The bottom of the screen has a button labeled ``END SESSION.” A scroll bar indicates continuation of the quiz.}
\end{figure*}

\begin{figure*}[hbt!]
\centering
\includegraphics[width=0.6\linewidth]{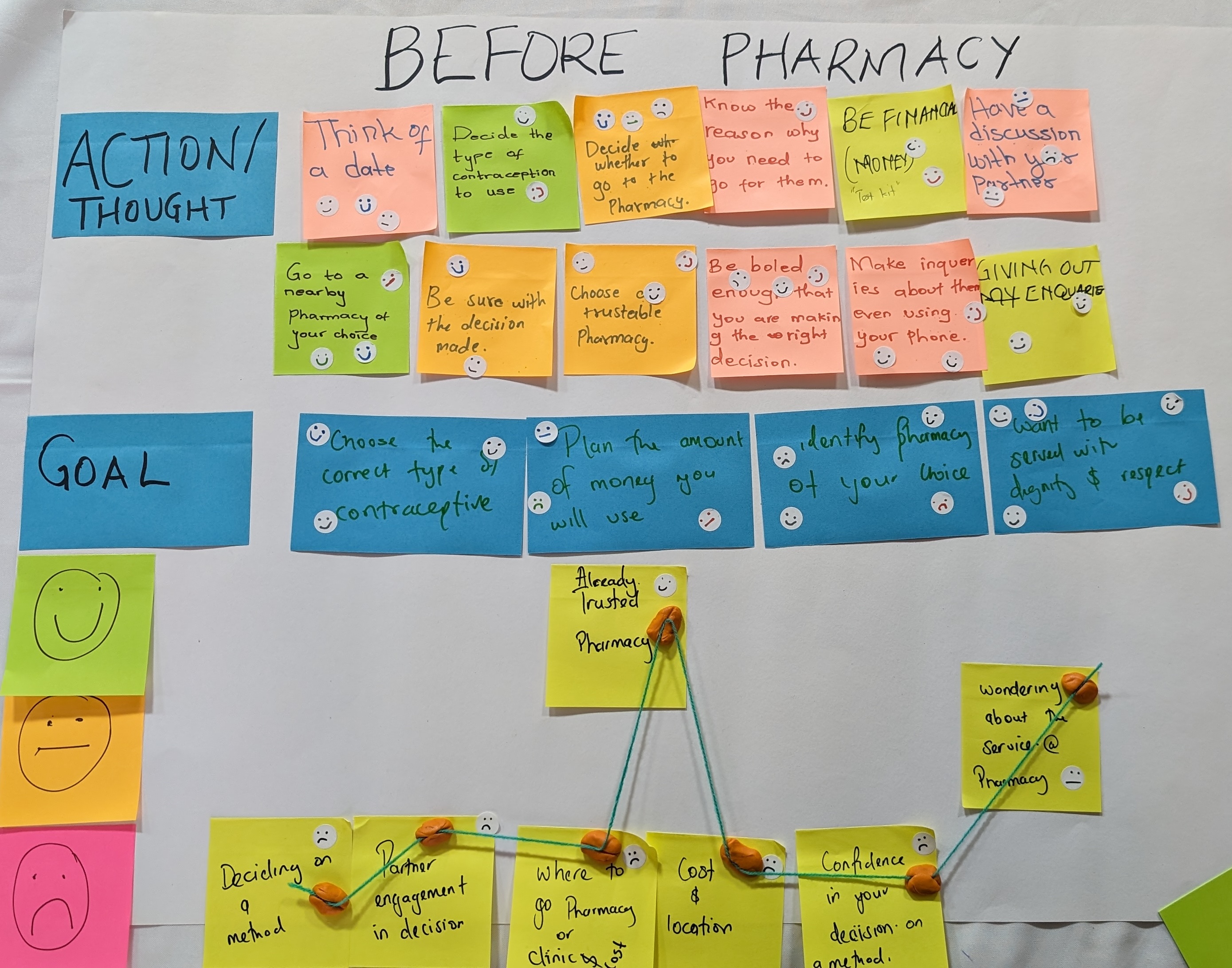}
\caption{AGYW journey mapping where AGYW and facilitators wrote actions, thoughts, goals, and emotions before a pharmacy visit on post-it notes.}
\Description{Poster paper with "BEFORE PHARMACY" at the top. Below, a post-it note labeled "ACTION/THOUGHT" and next to it multiple post-it notes with actions and thoughts AGYW have before visiting the pharmacy setting. Below is a post-it note labeled "GOAL" and next to it post-it notes of goals that AGYW have for visiting the pharmacy setting. Below are three post-it notes placed vertically, each with a different facial expression drawn on it, where one is happy, one is neutral, and one is sad. Post-it notes are placed next to it and connected with a string to indicate the emotional path. On each post-it note is a sticker with the same three facial expressions.}
\end{figure*}

\begin{figure*}[hbt!]
\centering
\includegraphics[width=0.6\linewidth]{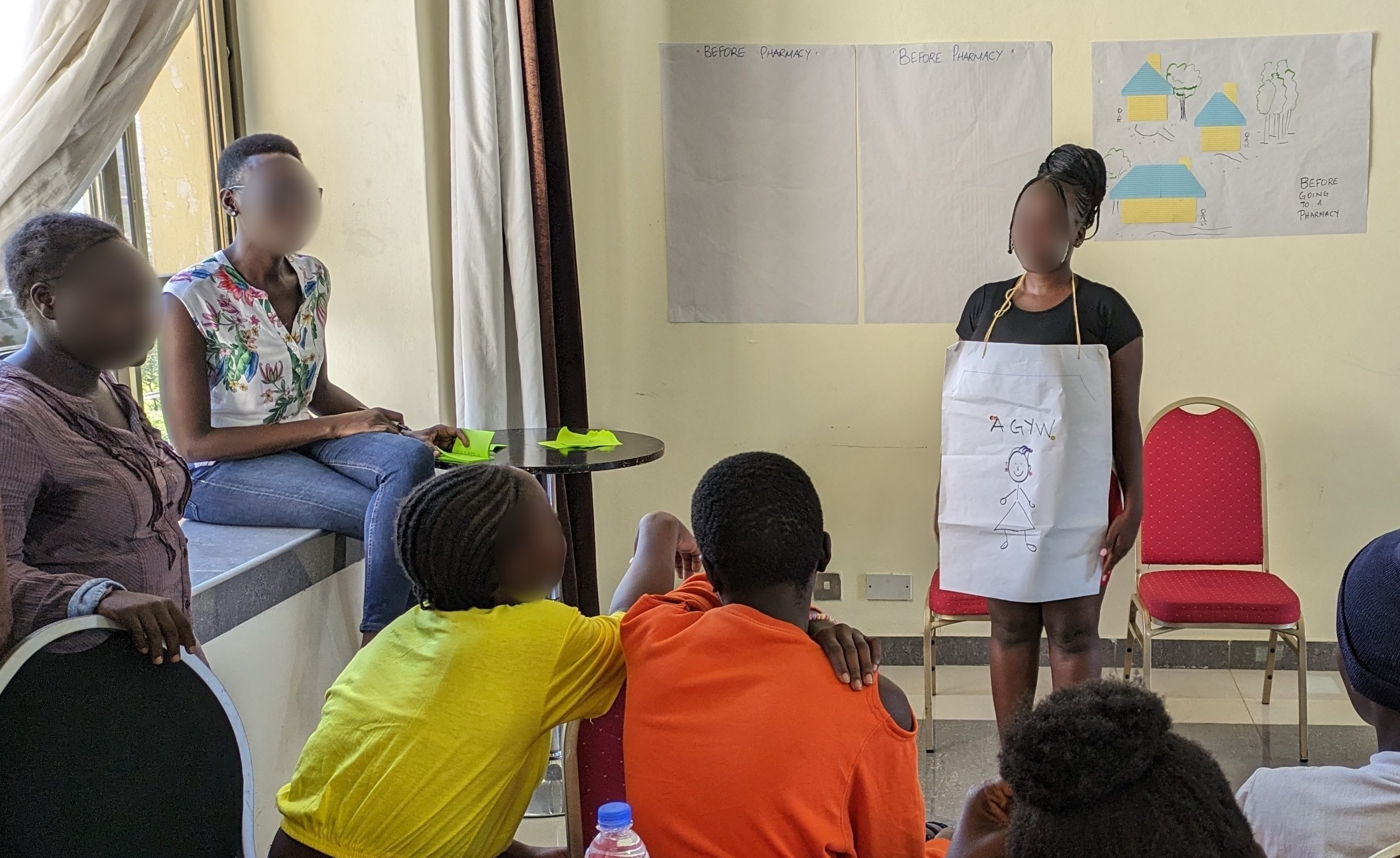}
\caption{AGYW journey mapping with role-play, led by the facilitator. }
\Description{A girl wearing a sign labeled "AGYW" with a drawing of a girl is facing five girls (workshop participants) and a woman (workshop facilitator).}
\end{figure*}

\begin{figure*}[hbt!]
\centering
\includegraphics[width=0.6\linewidth]{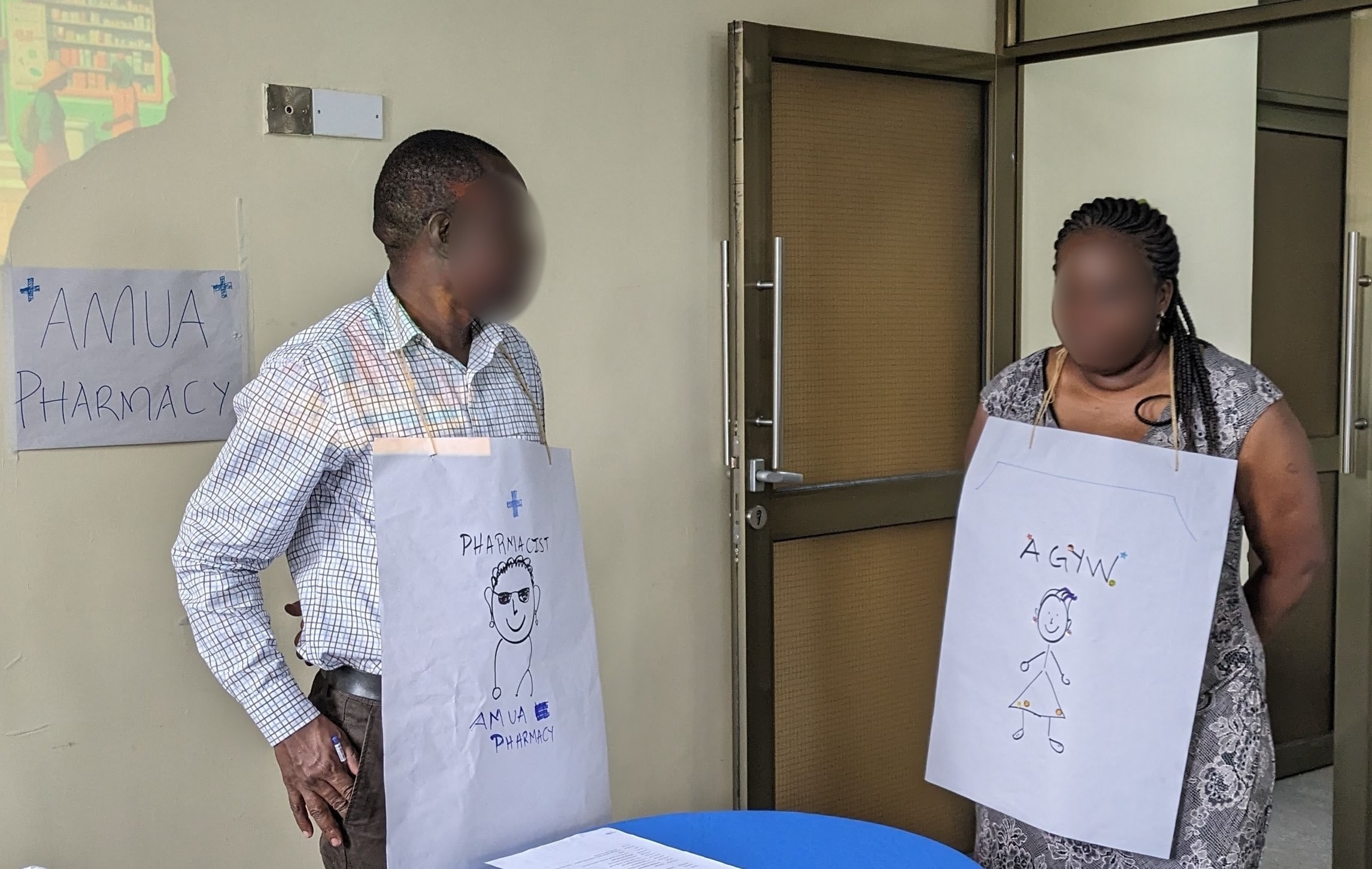}
\caption{Pharmacy staff journey mapping with role-play.}
\Description{A sign on the wall that says "AMUA PHARMACY". A man wearing a sign labeled "Pharmacist Amua Pharmacy" with a drawing of a man is talking to a woman with sign a labeled "AGYW" and a drawing of a girl.}
\end{figure*}

\begin{figure*}[hbt!]
\centering
\includegraphics[width=0.6\linewidth]{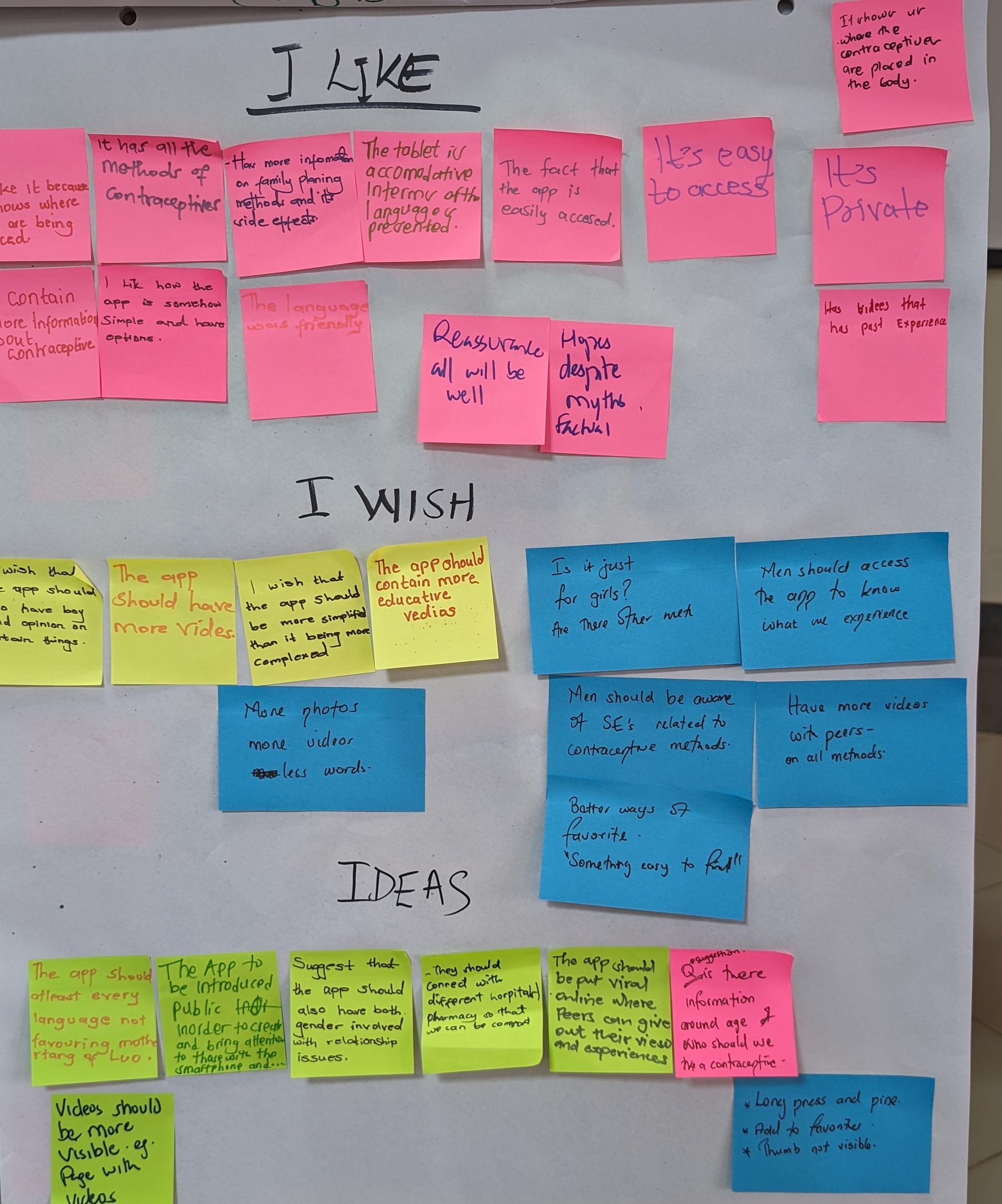}
\caption{App feedback documented by AGYW on post-it notes using the ``I Like, I Wish, What If” method.}
\Description{Poster paper with "I LIKE" at the top with post-it notes of comments on what participants liked about the app, "I WISH" with post-it notes with comments on  what participants wished the app had, and "IDEAS" with post-it notes with comments on ideas participants had with improving the app.}
\end{figure*}
